\documentclass[sigconf]{acmart}
\usepackage{tabularx}
\usepackage{booktabs}
\usepackage{enumitem}
\usepackage{graphicx}
\usepackage{fancyvrb}
\usepackage{xcolor}
\definecolor{LanguageRevision}{RGB}{32,55,100}
\newcommand{\langrev}[1]{#1}
\newcommand{\revone}[1]{#1}
\newcommand{\revtwo}[1]{#1}
\newcommand{\revoneR}[1]{#1}
\newcommand{\revtwoR}[1]{#1}
\newcommand{\revthree}[1]{#1}
\newcommand{\revonebody}[1]{#1}
\newcommand{\revtwobody}[1]{#1}

\newcommand{\revthreebody}[1]{#1}

\AtBeginDocument{%
  }

\copyrightyear{2026}
\acmYear{2026}
\setcopyright{cc}
\setcctype{by}
\acmConference[UIST '26]{The 39th Annual ACM Symposium on User Interface Software and Technology}{November 02--05, 2026}{Detroit, MI, USA}
\acmBooktitle{The 39th Annual ACM Symposium on User Interface Software and Technology (UIST '26), November 02--05, 2026, Detroit, MI, USA}
\acmDOI{10.1145/3830398.3830513}
\acmISBN{979-8-4007-2856-3/2026/11}

\begin{document}

\title{ASIDE: From Conflict Participants to Co-Observers Through Dyadic Spectator Reflection}

\author{Xinyi Zhang}
\orcid{0009-0009-1419-2772}
\affiliation{
\institution{School of Software Engineering\\Sun Yat-sen University}
\city{Guangzhou}
\country{China}}
\email{zhangxy2227@mail2.sysu.edu.cn}
\author{Jingting He}
\orcid{0009-0003-8720-6254}
\affiliation{
\institution{School of Software Engineering\\Sun Yat-sen University}
\city{Guangzhou}
\country{China}}
\email{hejt36@mail2.sysu.edu.cn}
\author{Zicheng Zhu}
\authornote{Corresponding author.}
\orcid{0000-0002-4332-2515}
\affiliation{
\institution{School of Computing\\National University of Singapore}
\city{Singapore}
\country{Singapore}}
\email{zicheng@u.nus.edu}
\author{Yuxin Su}
\orcid{0000-0002-3338-8561}
\affiliation{
\institution{School of Software Engineering\\Sun Yat-sen University}
\city{Guangzhou}
\country{China}}
\email{suyx35@mail.sysu.edu.cn}

\renewcommand{\shortauthors}{Zhang et al.}

\begin{abstract}
When two people argue over text, each knows what they meant and can only guess what the other was thinking. Existing AI reflection tools work from one person's account, and dyadic tools support co-expression without making the gap between accounts inspectable. We propose Dyadic Spectator Reflection (DSR), an interaction structure in which both partners externalize their models of each other independently and then encounter them together, and present ASIDE, a system that operationalizes it. ASIDE replays a past text conflict as a pixel-art theatrical scene where each character's unspoken state appears as an AI-inferred thought bubble either partner can contest and rewrite. Each edits alone, and the two versions meet only when both are done, in a scene they watch together. In an exploratory study, 10 couples revisited real conflicts and described the scene as a shared position from which to observe their own argument, stepping out of their roles without disengaging from it, and Divergence Cards as a way to locate specific interpretation gaps afterward. We contribute DSR as a reusable interaction structure, ASIDE as its system realization, and exploratory empirical findings on how couples used it.
\end{abstract}



\begin{teaserfigure}
  \includegraphics[width=\textwidth]{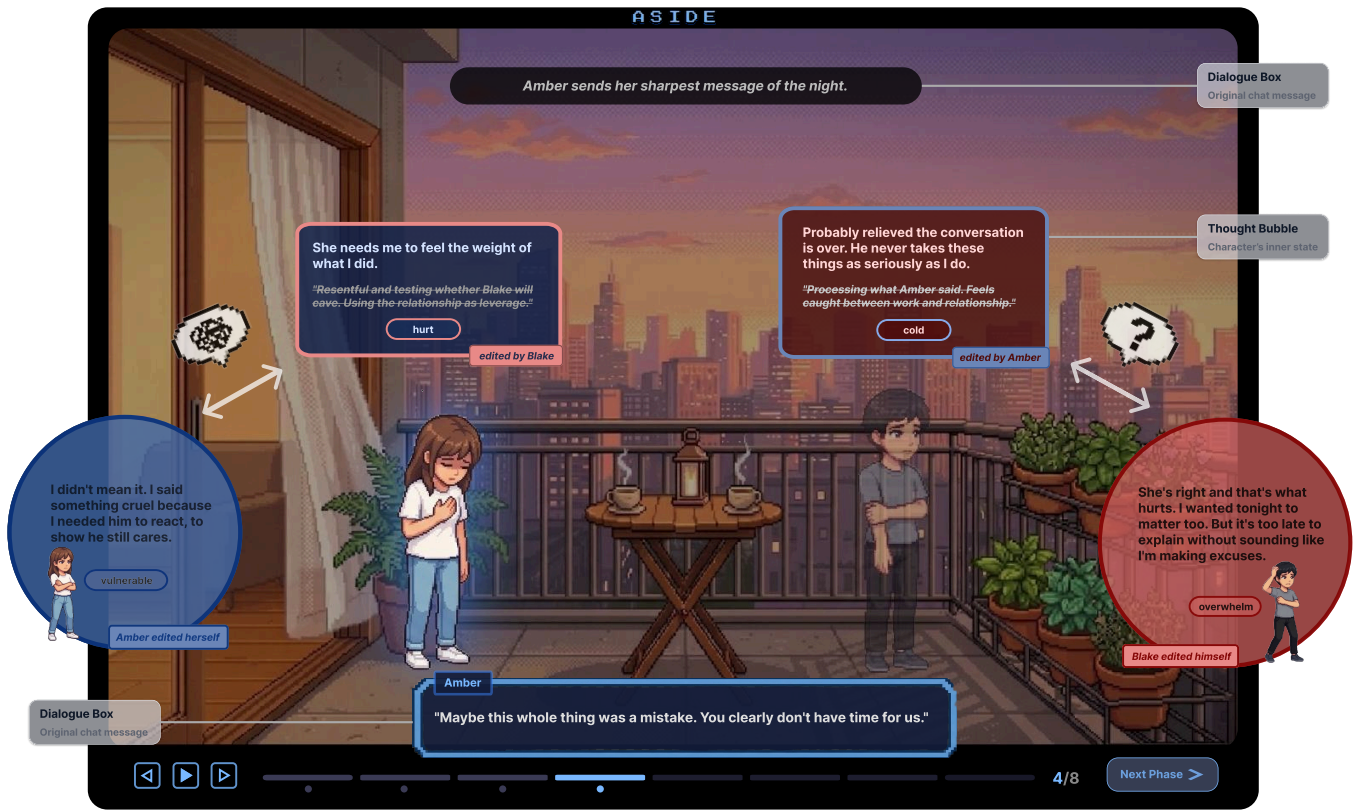}
  \caption{ASIDE reconstructs a couple's text conflict as a pixel-art theatrical scene. The center shows Together Viewing, where each thought bubble displays one partner's independently edited interpretation of the other's inner state. Struck-through text is the AI's original inference and text above is the partner's revision. Enlarged blue and red overlays show what Amber (left) and Blake (right) separately confirmed feeling across different ASIDE phases.}
  \Description{An annotated composite of one ASIDE theatrical beat.}
  \label{fig:teaser}
\end{teaserfigure}

\makeatletter
\pretocmd{\@mkabstract}{\clearpage}{}{}
\makeatother
\maketitle
\section{Introduction}
\label{sec:introduction}
After a text-based conflict, two people are left with the same chat log but different experiences of it~\cite{murray2002kindred}. Each carries a personal account of what happened, including what they tried to express, what they heard the other person say, and what they believed the other person was thinking. These accounts often diverge~\cite{sillars2000cognition,kenny2001accuracy}, yet neither person knows the extent of that divergence.
Text deepens this asymmetry because the chat log preserves sent messages while omitting hesitation, tone, and unsent revisions.
The resulting record feels complete even though important context is missing~\cite{culnan1987social,walther1996computer,byron2008carrying}.
Direct conversation offers a natural path to repair, but it requires enough emotional distance for both parties to listen with less defensiveness. It also requires a shared reference point that keeps the discussion from becoming a contest between two accounts~\cite{johnson2004practice}. These conditions are often absent after conflict, which makes renewed escalation more likely~\cite{gottman1999seven,hochschild1983managed,christensen1990}. Time alone may not restore them because unexamined attributions can consolidate and become harder to revisit~\cite{bradbury1990attributions,fincham2004forgiveness,karney2000attributions}.

People seek ways to step back before re-engaging. Revisiting personal records can support retrospective reflection, cognitively oriented conversations with close others can facilitate reappraisal and emotional recovery, and language models can offer alternative reframings of negative thoughts~\cite{isaacs2013echoes,nils2012beyond,sharma-etal-2023-cognitive}.
Written reflection on distressing experiences can also promote self-distancing and reduce later emotional reactivity~\cite{parkSteppingBackMove2016}.
Each approach still begins with one person's account, so the resulting insight reflects that person's framing. Dyadic tools involve both partners through shared journaling or co-reflection and make both experiences visible. However, side-by-side expression does not reveal what a partner felt behind a specific message or how one's own words shaped the partner's interpretation~\cite{10.1145/3706598.3713642,clark1996using}. Each person's inner experience therefore remains partly inaccessible. Perspective-taking alone also leaves this gap unresolved because it can increase confidence without improving accuracy~\cite{eyal2018mind,ickes2003everyday}. Mutual understanding requires each partner's experience to be externalized and open to correction.

\revtwo{We first conducted a formative interview study with 14 individuals in romantic relationships to understand how they reflected on text-based conflicts. The findings revealed barriers to articulating inner states, accessing the partner's perspective, and gaining distance from the conflict. We therefore asked how partners might step outside the conflict and view it together as co-observers. Drawing on prior work on scaffolding and self-distancing~\cite{wood1976tutoring,kross2011self,kross2017self}, we developed this direction into an interaction structure called \textbf{Dyadic Spectator Reflection (DSR)}.
DSR repositions both partners from defenders of separate accounts to co-observers of an interpersonal event. It frames mutual understanding as a shared reflective practice in which differing interpretations can be externalized, brought into view, and examined together.}

We present \textbf{ASIDE}, a system that operationalizes DSR.
ASIDE reconstructs a couple's text-based conflict as a pixel-art theatrical scene with two small figures and AI-inferred thought bubbles~\cite{park2023generative,10.1145/3772318.3790572,suh2024luminate}. The bubbles represent what each partner may have felt beneath the words they sent. Through a Cross-Perspective Editing Protocol, each partner confirms their own thoughts and independently edits the inferences about the other person. Once both partners finish, they view the co-annotated scene simultaneously and see how the other person interpreted them during the conflict.

\revone{We evaluated ASIDE with 10 couples (N=20) who reflected on past text-based conflicts. This exploratory study combines descriptive within-session measures, interaction logs, and interviews to characterize how participants used the workflow. Editing surfaced gaps between what participants had said and what they had felt, while shared viewing supported a shift toward co-observation.}

% This work contributes three outcomes. First, \textbf{DSR} offers an interaction structure through which both partners can observe a conflict together. Second, \textbf{ASIDE} operationalizes DSR through pixel-art theatrical reconstruction and a cross-perspective editing protocol. Third, \revone{\textbf{exploratory empirical findings} from 10 couples characterize how stepping outside a conflict together changes what partners notice and discuss.}

This work makes three contributions.
\begin{enumerate}
  \item \revtwo{DSR offers an interaction structure that positions partners as co-observers of a past conflict.}
  \item ASIDE operationalizes DSR through pixel-art theatrical reconstruction and cross-perspective editing.
  \item \revone{Exploratory findings from 10 couples characterize how partners used this structure and what it surfaced during shared reflection.}
\end{enumerate}

\section{Related Work}
\label{rw}

\subsection{The Participant Position in Intimate Conflict}
\label{rw:conflict}

\langrev{Interpersonal conflict is experienced from within each person's own role in the interaction. Studies of marital conflict show that overt talk captures only part of what participants think and feel, while attention and spontaneous attributions often center on a partner's intentions and the relationship implications of what was said~\cite{sillars2000cognition,sillars1998misunderstanding}. We use the term \textit{participant position} to describe this situated standpoint. Each person has first-hand access to their own experience but must infer the other's from incomplete behavioral evidence~\cite{ickes1993empathic,kenny2001accuracy}. These inferences can reflect assumed similarity as well as knowledge of the partner~\cite{kenny2001accuracy}, and current affect can constrain judgments about experiences outside one's present emotional state~\cite{loewenstein2005hot}. Negative sentiment and conflict-related attributions may further stabilize unfavorable interpretations over time~\cite{gottman1999seven,gottman2015principia,bradbury1990attributions,fincham2004forgiveness}. Partners can therefore leave the same exchange with little overlap between their experiences despite believing that they understand each other~\cite{sillars2000cognition}.}

\langrev{Perspective taking alone does not reliably overcome this asymmetry. People commonly begin from their own viewpoint and adjust insufficiently~\cite{epley2004perspective}; across 25 experiments, directly obtaining another person's perspective improved interpersonal accuracy whereas instructions to imagine that perspective did not~\cite{eyal2018mind}. Text communication adds another constraint because it removes gesture, emphasis, and intonation~\cite{noller1980misunderstandings,10.1145/2556288.2557177}. Senders consequently overestimate how accurately recipients will understand their intended tone~\cite{kruger2005email}. A chat log preserves the words exchanged, but not the unspoken concerns or intended tone needed to interpret them~\cite{sillars2000cognition,kruger2005email}. Self-distancing can change how a person relates to their own experience and reduce emotional reactivity~\cite{kross2011self,kross2017self}, but it cannot supply the partner's missing account. Dyadic reflection must therefore create a way for both accounts to be articulated and encountered together.}

\subsection{AI-Mediated Relationship Support}
\label{rw:mediated}

\langrev{Digital reflection tools have progressively expanded what users can capture, revisit, and reinterpret. Personal informatics systems make everyday experiences available for retrospective review~\cite{li2010stage,rooksby2014lived,epstein2015lived}. AI-assisted journals prompt users to articulate and reconsider emotional experiences~\cite{10.1145/3613904.3642693,nepal2024mindscape,isaacs2013echoes,kim2024mindfuldiary}, while editable AI drafts support iterative sensemaking in writing and creative work~\cite{10.1145/3491102.3502030,gero2023social,yuan2022wordcraft,chung2022talebrush}. DiaryPlay further transforms personal narratives into interactive scenes through which a viewer can role-play the author's perspective~\cite{10.1145/3772318.3790572}. In relationship-oriented settings, conversational systems provide emotional support or suggest alternative interpretations from the user's account~\cite{fitzpatrick2017delivering,skjuve2021my,sharma2023human}. These systems illustrate how AI can scaffold articulation and perspective exploration, but reflection still begins from one person's account.}

\langrev{LLM-based systems also support conflict preparation. Rehearsal lets users practice difficult conversations with a simulated interlocutor and receive strategy-based feedback~\cite{shaikh2024rehearsal,ebner2012games}, while ConflictLens analyzes records of real conflicts and provides exercises for reflection and communication practice~\cite{chun2025conflictlens}. Both ground useful reflection in concrete conflict material, yet the other party's experience is represented through one user's account or an AI simulation rather than through that person's direct participation.}

\langrev{Other work brings technology into communication between people. AI-mediated communication systems modify, augment, or generate messages on behalf of a communicator~\cite{hancock2020aimc}. Design research on hard conversations identifies opportunities to support mutual consent, emotional regulation, and well-timed pauses during digitally mediated conflict~\cite{baughan2024hardconversations}. Dyadic systems move further toward shared reflection. In TogetherReflect, partners separately visualize emotions, enter each other's drawings for discussion, and then create a shared canvas~\cite{10.1145/3706598.3713642}. This work demonstrates the value of combining individual expression with joint engagement. However, it does not focus on comparing how both partners interpreted the same moments in an interaction record. Prior work therefore leaves underexplored a retrospective structure in which partners independently revise interpretations of a shared conflict and then examine where those interpretations diverge.}

\subsection{Externalization and Shared Observer Perspectives}
\label{rw:externalization}

\langrev{Externalization turns tacit knowledge into a representation that can be inspected and revised~\cite{schon1983reflective,sengers2005reflective}. Once an external representation becomes part of a group's shared context, its constraints and visual salience can guide what collaborators notice and discuss~\cite{suthers2001towards}. Shared artifacts can also coordinate people who hold different viewpoints without requiring them to collapse those viewpoints into a single consensus~\cite{star1989institutional}. Studies of shared workspaces and prototypes similarly show how external artifacts organize attention and interaction during collaborative work~\cite{brereton2000observational}. HCI systems operationalize these ideas in tools such as Sensecape, which organizes generated information for visual exploration~\cite{suh2023sensecape}, and Graphologue, which turns conversational content into editable graphs~\cite{jiang2023graphologue}. These systems demonstrate the value of inspectable representations, although their artifacts primarily support information-focused sensemaking or task coordination rather than comparison of subjective accounts in an interpersonal conflict.}

\langrev{Representation also shapes the standpoint from which an experience is considered. Self-distancing research shows that adopting a third-person perspective can reduce emotional reactivity while preserving attention to the meaning of an experience~\cite{kross2011self,kross2017self,ayduk2010distance,grossmann2014solomon}. Construal-level research similarly links psychological distance with more abstract processing~\cite{trope2010construal}. These findings concern an individual's perspective on personal experience; they do not by themselves establish a shared observer position for two people. DSR brings these strands together. It externalizes each partner's account, preserves differences between the two accounts in a shared artifact, and uses theatrical abstraction to help partners encounter the conflict as co-observers. The aim is to make subjective differences inspectable without presenting any account as a single objective truth.}

\section{Design}
\label{sec:design}

\revthree{ASIDE emerged from an iterative process that combined a formative study, concept development, and prototype testing. This section presents the observations that motivated the design~(\S\ref{sec:formative}), defines Dyadic Spectator Reflection (DSR) and its three design principles~(\S\ref{sec:dp}), and describes the iterations that produced the current system~(\S\ref{sec:iterations}).}

\subsection{Formative Study}
\label{sec:formative}

\subsubsection{Survey (N=90)}

An online survey of young adults in romantic relationships (ages 18--36; $M=22.33$, $SD=3.25$) examined their use of AI after relationship conflict. Among these users, 64.4\% sought emotional relief, 59.3\% used AI to analyze what had happened, 59.3\% talked through the conflict with AI, and 55.9\% used it to infer the other person's actions or thoughts. Respondents could select multiple purposes. Mean trust in AI-generated relational content was 5.51/7 ($SD=0.84$). Participants neither rejected AI-inferred content outright nor accepted it uncritically, which supported our use of editable AI scaffolds.

\subsubsection{Interviews (N=14)}

We conducted semi-structured interviews lasting 30--40 minutes with 14 individuals (8F, 6M, ages 19--28) in romantic relationships. All had experienced a text-based conflict and used an AI tool in response during the previous month. Three observations emerged.

\paragraph{O1. Reflection needs a starting point without a verdict.}
Without external structure, post-conflict reflection rarely begins on its own. P1 noted that ``initiating a topic costs energy'' and that ``once the emotional moment passes, you can never recover that same intensity later.'' P11 wanted something ``like a diary or memo that captures what happened without me having to deliberately press something.'' Such a structure could lower the effort required to revisit the conflict.

Participants also distinguished an AI thinking aid from a relational authority. P4 found AI feedback ``very fair'' and felt less sad, while P5 said it ``eliminates many unnecessary misunderstandings.'' P9 still demanded ``absolute objectivity'' and did not want the relationship to rely on AI. P12 expected real-time intervention during conflict to ``backfire,'' although she considered post-hoc reflection less intrusive. P8 summarized the desired boundary by saying, ``It doesn't need to do anything for you. It just needs to show you the other person's state.'' This pattern aligns with evidence that LLMs can affirm the perspective presented by a user and leave problematic assumptions unchallenged~\cite{sharma2024towards,cuadra2024illusion}. Reflection therefore needs an external prompt that remains open to correction. This observation motivated \emph{DP1}.

\paragraph{O2. Understanding requires both voices.}
Solo AI interaction captures only one side of a conflict. P2 warned that separate AI consultations produce ``two AIs analyzing the conflict rather than two people understanding each other.'' Her partner's AI had ``kept exaggerating and deepening the conflict'' after hearing only his account. Baughan et al.~\cite{baughan2024hardconversations} similarly found that digital tools for close-relationship conflict often lack mechanisms for mutual consent to enter reflective space. P3 described the emotional cost of this asymmetry and said that using AI alone made her feel like ``the only one putting in effort for our emotions.''

Participants also described a deadlock around initiating joint reflection. P6 had no way to know whether his partner had calmed down, and P12 responded to conflict by becoming silent even though she knew this harmed the relationship. P3 observed that unresolved conflicts become more volatile over time. These accounts reveal two barriers. Partners cannot signal readiness simultaneously without one person bearing the cost of reopening the issue, and existing tools do not capture both accounts of the conflict. This observation motivated \emph{DP2}.

\paragraph{O3. Reflection needs distance and emotional texture.}
Participants wanted to step outside the conflict and examine it from another position. P4 wanted a ``God's-eye view'' and P10 wanted an observer's perspective on the partner's subtext. These accounts point to a need for representations that help partners observe a conflict without immediately re-entering it.

Participants also wanted the representation to retain enough emotional detail for meaningful reflection.
P14 found their responses generic and preferred a sequence in which the AI first acknowledged emotion and introduced the other perspective after the user had calmed down. P3 described existing AI personas as ``stiff, particularly cold,'' while P10 felt that the output lacked a natural human quality.
P11 added that text records lose the vivid emotional context of the original moment. These experiences align with critiques of shallow emotional engagement in LLM-generated content~\cite{cuadra2024illusion,sharma2023human}. The need to combine analytical distance with emotional texture motivated \emph{DP3}.

\subsection{Design Principles}
\label{sec:dp}

\revthree{Our design goal was to help partners move from defending separate accounts to examining a shared representation as co-observers. We call this shift \textbf{Dyadic Spectator Reflection (DSR)} and describe its sequence as \textit{externalize independently, then encounter together}. Three design principles (DPs) operationalize this sequence. Together, the principles make AI inferences revisable, preserve independent contributions before reveal, and support reflective distance during shared viewing.}

\paragraph{DP1. Scaffold, Don't Solve}
\label{sec:dp1}

\revthreebody{Articulating inner states after conflict presents a double bind. An empty page offers too little structure, while a definitive AI analysis can create false certainty. Eyal et al.~\cite{eyal2018mind} found that people stop examining a partner's perspective when they believe they already understand it. An AI-generated interpretation should be concrete enough for users to assess while remaining provisional enough to invite correction.}

\revthreebody{ASIDE provides an \emph{imperfect scaffold}~\cite{wood1976tutoring}. For each conflict beat, the system generates a tentative inference about each person's inner state. Users can accept, modify, or replace every inference, and acceptance requires the same deliberate action as revision. This design turns an open-ended recall task into a directed judgment about whether a proposed interpretation fits. When the AI's reading differs from the user's experience, the discrepancy gives the user a concrete interpretation to correct and may help them articulate what they actually felt.}

\paragraph{DP2. Symmetric Encounter}
\label{sec:dp2}

\revthreebody{The empathy gap persists when each partner’s interpretation of the other’s internal state remains invisible ~\cite{loewenstein2005hot}. Making annotations visible in real time, however, may allow one partner’s responses to shape the other’s before both accounts are independently articulated ~\cite{ross1991reactive,hancock2020aimc}. A symmetric encounter should therefore preserve independent expression before making both perspectives mutually visible.

ASIDE therefore asks partners to \emph{edit separately, then see together}. 
Each partner first develops their own account. Once both are complete, the system reveals them simultaneously. This sequence limits early mutual influence and turns differences between the accounts into shared objects for inspection and discussion.}

\paragraph{DP3. Distance Without Disconnection}
\label{sec:dp3}

\revthreebody{Revisiting a conflict requires a careful balance. Strong immersion can reactivate defensive responses, while excessive abstraction can remove the emotional relevance that makes reflection worthwhile. Research on self-distancing shows that a third-person perspective can reduce emotional reactivity while preserving analytical depth~\cite{kross2011self,kross2014self}. Negative sentiment override can also lead conflict participants to assign negative intent to ambiguous behavior~\cite{gottman1999seven}.
A reflective representation should create enough distance while retaining enough emotional detail to keep the conflict recognizable and relevant.}

\revoneR{ASIDE uses pixel-art theatrical abstraction to balance involvement and distance. The characters represent the partners in their original roles. We match gender for recognizability while stylizing age and appearance to avoid overly literal re-immersion. The scene places an exchange that may have unfolded across locations into a single shared setting. By shifting ``me and you arguing'' into ``two small figures on a stage,'' it keeps the conflict recognizable while supporting a co-observer stance.}

\subsection{Design Iterations}
\label{sec:iterations}

We refined ASIDE through three pilot iterations involving three couples. Observed interaction breakdowns guided the design changes described below.

\paragraph{Iteration 1. Theatrical staging.}
An initial text-only format kept users immersed in the original exchange. Pixel-art theatrical staging encouraged third-person language and an observer stance. Participants found realistic visuals ``embarrassing,'' while pixel art felt ``playful and comfortable'' and still recognizable as ``about us''.

\paragraph{Iteration 2. Self-first ordering.}
We explored a partner-first sequence in which users interpreted the partner's inner states before reflecting on their own. Two order effects emerged. Participants sometimes projected their own feelings onto the partner, as one explained, ``I was writing what I felt, not what he felt.'' Considering the partner's perspective could also soften participants' emotions before self-confirmation, making their later account less representative of how they had felt during the conflict. These effects reflect known limits of introspection and perspective-taking~\cite{nisbett1977telling,eyal2018mind}. ASIDE therefore asks each partner to articulate their own experience before interpreting the other's.

\begin{figure}[t]
  \centering
  \includegraphics[width=\columnwidth]{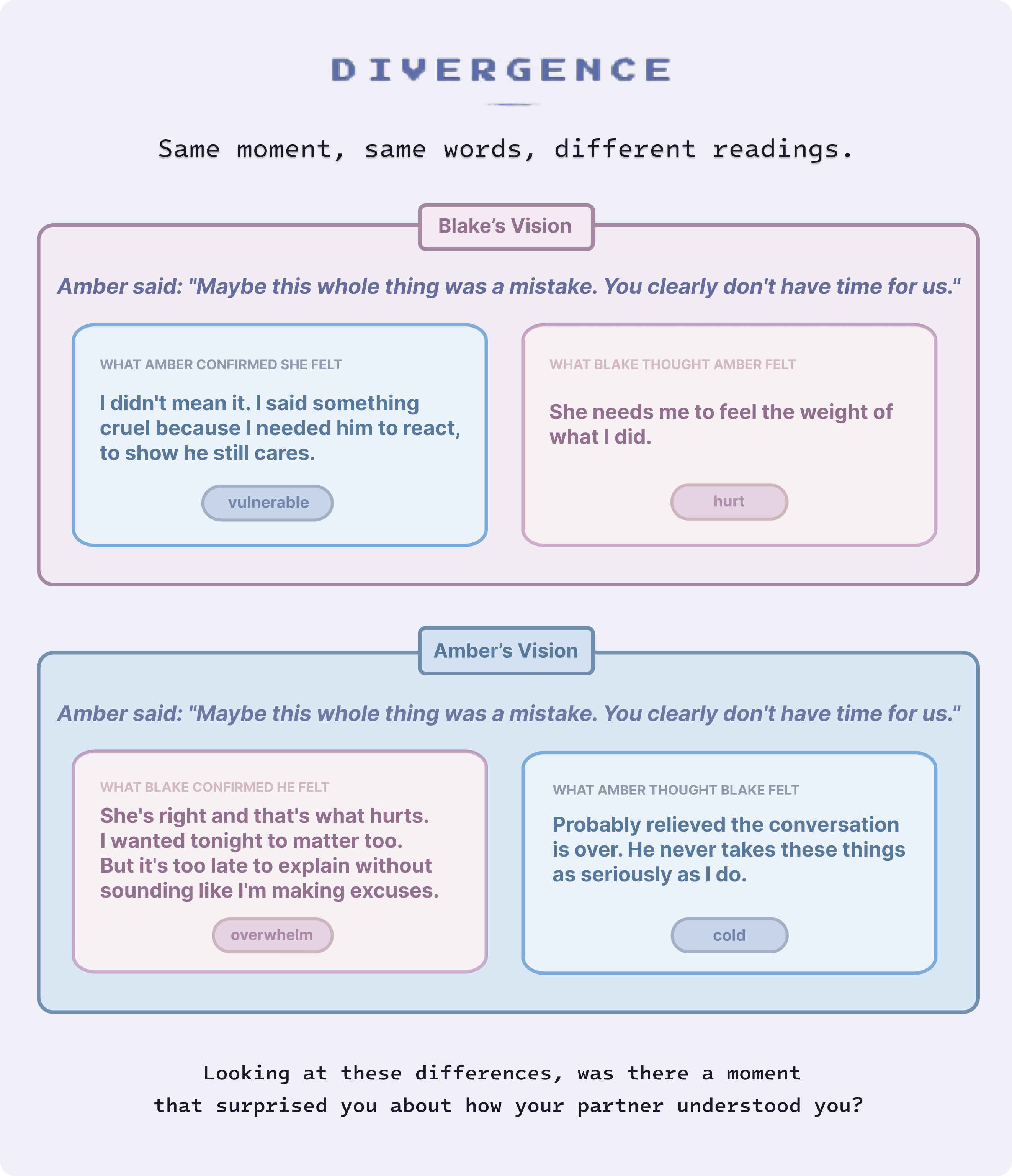}
  \caption{Divergence cards compare two readings of the same conversational moment. Each card pairs one person's confirmed feelings with the partner's interpretation. In this beat, Amber felt \emph{vulnerable}, while Blake interpreted her as trying to \emph{hurt} him. Blake felt \emph{overwhelmed}, while Amber interpreted him as \emph{cold}.}
  \Description{Two Divergence Cards compare partners' interpretations of the same conversational beat. The upper card places Amber's self-confirmed feeling of vulnerability beside Blake's interpretation that she wanted him to feel hurt. The lower card places Blake's self-confirmed feeling of being overwhelmed beside Amber's interpretation that he felt cold.}
  \label{fig:divergence-cards}
\end{figure}

\paragraph{Iteration 3. Divergence cards.}
Without a structured artifact, discussion after viewing returned to explanations of each person's position. We therefore added a beat-level comparison that placed one partner's confirmed inner state beside the other's interpretation of that state. Making these differences visible helped partners move from recognizing disagreement to examining why specific interpretations diverged. We implemented this design as the Divergence Cards shown in Figure~\ref{fig:divergence-cards}.
\revtwoR{The scene and cards therefore have complementary roles. The theatrical scene provides a shared space for embodied, emotionally distanced co-observation. The cards provide post-viewing comparison artifacts that make interpretation gaps easier to locate, revisit, and discuss.}

\begin{figure}[t]
  \centering
  \includegraphics[width=\columnwidth]{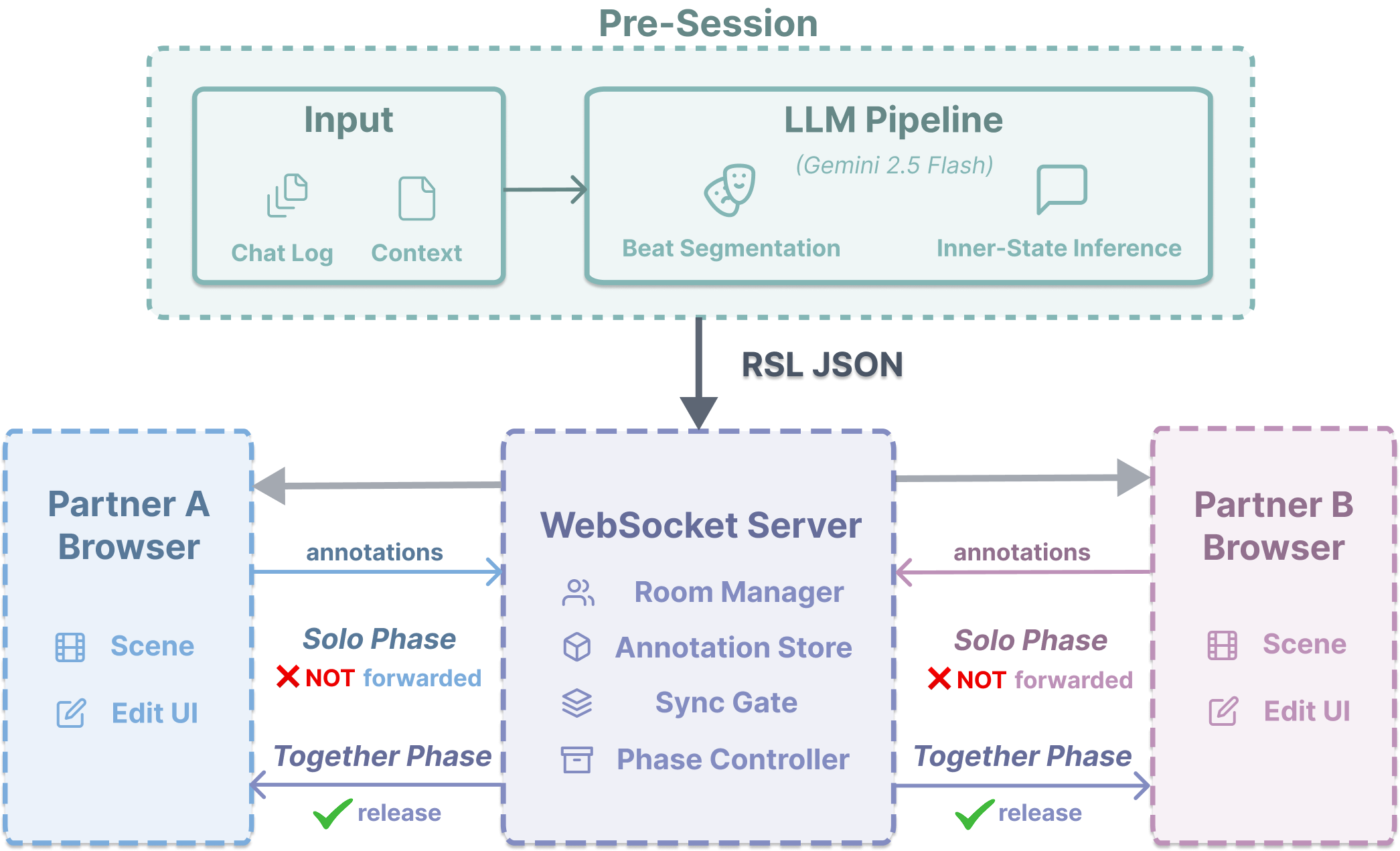}
  \caption{ASIDE's architecture includes a pre-session reconstruction pipeline, a central WebSocket server, and separate browser clients for Partners A and B. The pipeline transforms the chat log and contextual input into an RSL scenario.}
  \Description{ASIDE's architecture includes a pre-session reconstruction pipeline, a central WebSocket server, and separate browser clients for Partners A and B. The pipeline transforms the chat log and contextual input into an RSL scenario. During private editing, each browser sends annotations to the server without forwarding them to the other partner. During Together Viewing, the server releases the completed annotations and synchronizes both clients.}
  \label{fig:architecture}
\end{figure}

\section{System}
\label{sec:system}

\subsection{System Overview}
\label{sec:architecture}
ASIDE is a web application for two partners to revisit a past text-based conflict. Each partner joins through a separate browser session connected to a shared backend. The interface reconstructs the original conversation as a sequence of pixel-art theatrical beats. It preserves the dialogue while presenting revisable inner-state hypotheses in thought bubbles above the two characters (Figure~\ref{fig:teaser}).

\revthree{Figure~\ref{fig:architecture} shows how ASIDE's pre-session reconstruction pipeline, two browser clients, and WebSocket synchronization server work together. The reconstruction pipeline transforms the chat log into structured RSL data. Each browser renders the scene and supports annotation, while the server controls information visibility, stores annotations, and coordinates the shared reveal.}

\begin{figure}[t]
  \centering
  \includegraphics[width=\columnwidth]{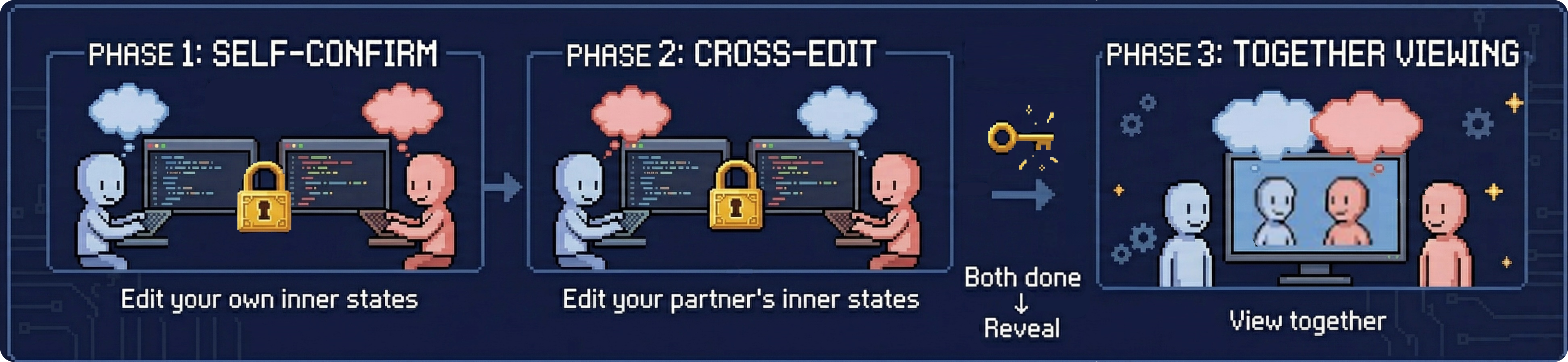}
  \caption{ASIDE's workflow from private reconstruction to shared spectatorship.}
  \Description{A three-phase workflow illustrated with blue and pink characters. The partners first edit their own inferred inner states separately, then edit their interpretations of each other while their work remains hidden. A synchronization gate opens after both finish, allowing them to view the reconstructed conflict together.}
  \label{fig:workflow}
\end{figure}

\revthree{Together, these components support a three-phase interaction workflow.
Figure~\ref{fig:workflow} summarizes how ASIDE realizes DSR's sequence of \emph{externalize independently, then encounter together}. During \emph{Self-Confirm}, each partner reviews and corrects AI-generated hypotheses about their own inner states. During \emph{Cross-Editing}, each independently reviews and edits hypotheses about the other partner's inner states. Each partner's annotations remain hidden until both finish. During \emph{Together Viewing}, the system reveals the completed annotations simultaneously and synchronizes the theatrical replay for joint viewing.
After the replay, partners review Divergence Cards that compare confirmed and partner-interpreted inner states at specific beats (Figure~\ref{fig:divergence-cards}).}

\subsection{Conflict Reconstruction}
\label{sec:pipeline}

\revthree{Before annotation begins, the reconstruction pipeline uses Gemini~2.5~Flash to transform the chat log and participants' contextual input into a structured RSL scenario. The model selects representative dialogue beats, generates revisable inner-state hypotheses, and specifies scene and staging metadata.}

\paragraph{Beat segmentation.}
The system divides the conversation into \emph{beats}, each of which contains one or two representative messages. For longer conversations, it selects escalation points, key misunderstandings, failed attempts at de-escalation, and gaps in communication. This sequence follows the conflict arc described by Gottman~\cite{gottman1999seven} and usually contains 5--12 beats. The original dialogue remains verbatim and locked. Users can edit the interpretation of what was felt, but they cannot alter what was said. This separation keeps the dialogue as common factual ground while allowing interpretations to diverge~\cite{clark1991grounding}.

\begin{table}[t]
  \caption{Information visibility and editing targets across the Cross-Perspective Editing Protocol.}
  \label{tab:visibility}
  \small
  \begin{tabular}{@{}lccl@{}}
    \toprule
    \textbf{Phase} & \textbf{A sees} & \textbf{B sees} & \textbf{Sync} \\
    \midrule
    Self-Confirm   & A's bubbles & B's bubbles & Independent \\
    Cross-Editing  & B's bubbles & A's bubbles & Independent \\
    \midrule
    \multicolumn{4}{c}{\emph{\small Both partners complete editing before release}} \\
    \midrule
    Together Viewing & \multicolumn{2}{c}{A (by B) + B (by A)} & Synchronized \\
    \bottomrule
  \end{tabular}
  \par\smallskip
  {\footnotesize
  During Together Viewing, ``A (by B)'' means A's inner state as edited by B, and ``B (by A)'' means B's inner state as edited by A. The interface also shows the original AI inference in smaller struck-through text.}
\end{table}

\paragraph{Inner-state inference.}
Before entering the main workflow, each partner independently completes a brief intake page describing their central concern, dominant emotion, and relevant situational context. Gemini~2.5~Flash uses these inputs together with the full chat log to generate a revisable inner-state hypothesis for each partner at every beat. Each hypothesis includes first-person inner-state text, an emotion tag, and corresponding pose and spatial metadata. The prompt requires the original dialogue to remain verbatim and instructs the model to avoid unsupported events or exaggerated emotional interpretations. The results are returned as structured RSL data (Appendix~\ref{app:rsl}). Appendix~\ref{app:prompts} documents the generation settings and prompt constraints~\cite{kim2024mindfuldiary,suh2024luminate}.

\paragraph{Scene rendering.}
\revthree{The LLM does not generate bitmap or pixel-art images. It produces structured beat specifications from the chat context. The frontend then selects a context-matched background from a pre-generated library of 25 pixel-art scenes and renders pre-made characters, props, thought bubbles, layout, and tension cues in real time.} Emotion tags control character poses and expressions~\cite{park2023generative,mccloud1993understanding}. Stage position reflects the proxemic state of the conflict~\cite{hall1966hidden}, with characters moving closer during moments of connection and farther apart during escalation. Facing direction indicates engagement or withdrawal. A factual narrator line supplies situational context without interpreting emotion. Section~\ref{sec:dp3} explains the pixel-art rationale, and Appendix~\ref{app:rendering} provides rendering details.

\subsection{Annotation and Synchronization}
\label{sec:dyadic}

The annotation interface controls what each user can do, and the synchronization protocol controls when those actions become visible to the partner. Together, they preserve independent editing before the shared reveal.

\paragraph{Annotation interface.}
Each AI-inferred inner state appears in a thought bubble above a character. Users can accept the inference, edit its text, or replace it.
An optional emotion label can also be accepted, changed, or dismissed.
Unconfirmed inferences receive no visual badge, and users may leave a beat unresolved when neither the inference nor an alternative feels appropriate. During annotation, users can manually navigate between beats and revisit earlier inferences as needed.

\paragraph{Synchronization protocol.}
The server manages WebSocket rooms for each pair. During Self-Confirm and Cross-Editing, it stores each partner's edits while hiding the text and progress from the other person. After both partners mark editing as complete, the server releases both sets simultaneously~\cite{ross1991reactive}. Together Viewing also synchronizes beat navigation so that both partners see the same moment~\cite{johansen1988groupware}. Table~\ref{tab:visibility} summarizes visibility across phases.
\revthree{Appendix~\ref{app:sync} documents the message protocol, phase-gate mechanism, annotation exchange, and behavioral event logging.}

\section{User Study}
\label{sec:study}

\revone{We conducted an exploratory study with 10 couples ($N=20$) who used ASIDE to revisit a past text-based conflict.} \revtwo{We examined how Dyadic Spectator Reflection unfolded as partners first externalized their interpretations independently and then encountered both accounts together. Two research questions (RQs) guided the study.}

\textbf{RQ1.} What do participants discover about themselves and each other through editing AI-inferred inner states?

\textbf{RQ2.} What happens when both partners' edited perspectives are revealed in the shared theatrical frame?

\subsection{Participants}
\label{sec:participants}

\revtwo{We focused this study on young-adult couples because relationship conflict commonly occurs through text in this population and leaves authentic chat records for examining dyadic reflection~\cite{leonard2025navigating}.} We recruited participants through an online screening survey distributed through university social media and personal networks. Both members of a couple had to agree to participate. Eligible couples needed an ongoing relationship and a preserved chat log from a text-based conflict that had occurred more than one week earlier. 
\revoneR{The one-week interval allowed time for ordinary reflection or repair before we examined what further perspectives emerged during ASIDE use~\cite{bradbury1990attributions,fincham2004forgiveness}.}
\revtwo{We excluded relationships involving intimate partner violence or severe psychological distress because synchronized disclosure may create unacceptable risks without clinical or crisis support. Each partner provided informed consent independently, and withdrawal by either person ended the session for the pair. Each participant received US\$8 in compensation for completing the study.}

The final sample comprised 10 couples ($N=20$). Participants had a mean age of 22.9 years ($SD=4.3$, range 18--36), and relationship duration ranged from less than six months to more than three years. The sample included 8 mixed-sex couples, 1 male--male couple, and 1 female--female couple. Of the 20 participants, 13 (65\%) had used AI tools after conflicts. Participant labels follow the format P\textit{n}a/P\textit{n}b. Table~\ref{tab:participants} summarizes participant demographics and prior AI use.

\begin{table}[t]
  \caption{Participant demographics, relationship duration, and prior AI use after conflict.}
  \label{tab:participants}
  \small
  \begin{tabular}{@{} c c c c c @{}}
    \toprule
    \textbf{Pair} & \textbf{Age} & \textbf{Gender}
      & \textbf{Duration} & \textbf{AI} \\
    \midrule
    P1  & 23, 21 & F, M & 2--3 yr    & No, No  \\
    P2  & 20, 20 & M, F & 6mo--1yr   & Yes, No \\
    P3  & 23, 23 & F, M & 2--3 yr    & Yes, No \\
    P4  & 19, 19 & F, M & 6mo--1yr   & Yes, Yes \\
    P5  & 23, 28 & M, M & $<$6 mo    & Yes, No \\
    P6  & 20, 20 & F, M & 1--2 yr    & No, No  \\
    P7  & 18, 19 & F, M & $>$3 yr    & Yes, Yes \\
    P8  & 23, 25 & F, F & 6mo--1yr   & Yes, Yes \\
    P9  & 26, 23 & F, M & $<$6 mo    & Yes, Yes \\
    P10 & 29, 36 & F, M & $>$3 yr    & Yes, Yes \\
    \bottomrule
  \end{tabular}
\end{table}

\subsection{Procedure}
\label{sec:procedure}

All sessions were conducted remotely by video conference, with each partner joining from a private location. Each session lasted approximately 60 minutes. Both partners first reread the original chat log independently and completed the PRE questionnaire. Each then provided their central concern, dominant emotion, and relevant situational context through an intake page. They subsequently reviewed and edited AI-generated hypotheses about their own inner states during Self-Confirm. During Cross-Editing, each person independently edited the inferences about their partner. Communication and progress visibility remained disabled during both phases.
After both partners completed editing, a synchronization gate released them into Together Viewing at the same time.
Voice and video were restored, and the partners first watched the composite scene together. 
\revoneR{After viewing, they reviewed Divergence Cards for relevant beats and discussed the highlighted interpretation gaps.}
Participants then completed the POST questionnaire independently and joined a semi-structured interview. 

\revtwo{A two-tier de-escalation protocol specified a five-minute cooling-off pause for lower-level escalation and session termination with support resources for more serious escalation~\cite{baughan2024hardconversations}. No intervention was triggered in the 10 study sessions.} 
\revthree{Sessions, interviews, and questionnaires were conducted in Chinese, the participants' native language. Original chat logs remained untranslated during system use, and the LLM generated inner-state hypotheses in natural Chinese. We translated the participant quotations and questionnaire items reported in the paper into English.} \revtwo{Before analysis, we replaced identifying information in interview transcripts and chat logs with pseudonyms and processed the research data locally.}

\subsection{Measures and Analysis}
\label{sec:measures}

\revtwoR{Our analysis centered on qualitative interviews to understand how participants experienced and interpreted the reflection process, while behavioral logs captured corresponding interaction patterns and questionnaire results provided complementary session-level context~\cite{creswell2017research}.}

\paragraph{Questionnaire scales.}
\label{sec:scales}
PRE and POST questionnaires used 7-point Likert scales from 1 (Strongly Disagree) to 7 (Strongly Agree). POST repeated the four PRE scales and added three items measuring scaffolding experiences. Table~\ref{tab:scales} summarizes the scales, sources, and research-question mappings, while Appendix~\ref{app:questionnaire} lists the full items.
PRE serves as a pre-use baseline for characterizing changes observed during the ASIDE session.

\revone{For each scale, we averaged its item scores to calculate one PRE and one POST score per participant. The negatively worded item in S1 was reverse-scored before averaging. We assessed the normality of participant-level change scores using Shapiro--Wilk tests. S1 and S2 were analyzed using two-sided paired-samples $t$-tests, while S3 and S4 were analyzed using two-sided Wilcoxon signed-rank tests because their change scores were non-normal. We report Cohen's $d_z$ for paired $t$-tests and rank-biserial correlation $r_{\mathrm{rb}}$ for Wilcoxon tests, together with medians and individual change directions. Because partners within a couple may not provide independent observations, we also averaged the two partners' change scores within each couple and conducted two-sided one-sample $t$-tests on the resulting 10 dyad-level changes as a sensitivity analysis. All tests used $\alpha=.05$. We interpret the unadjusted results as exploratory within-session patterns rather than causal efficacy evidence. POST-only S5 items are reported descriptively.}

\begin{table}[t]
  \caption{Questionnaire scales. S1--S4 were administered at PRE and POST. S5 was administered only at POST.}
  \label{tab:scales}
  \small
  \begin{tabular}{@{} l l c l l @{}}
    \toprule
    \textbf{Scale} & \textbf{Construct} & \textbf{Items}
      & \textbf{Source} & \textbf{RQ} \\
    \midrule
    S1 & Perspective-Taking     & 4 & IRI-PT~\cite{davis1983}& RQ1,2 \\
    S2 & Perspective Confidence & 3
      & Self-designed\textsuperscript{a}
      & RQ1,2\\
    S3 & Conflict Engagement    & 4
      & RRQ~\cite{trapnell1999}\textsuperscript{b} & RQ2 \\
    S4 & Empathic Concern       & 3 & IRI-EC~\cite{davis1983}
      & RQ1,2 \\
    \midrule
    S5 & Scaffolding Effect.    & 3
      & Self-designed\textsuperscript{c} & RQ1 \\
    \bottomrule
  \end{tabular}

  \raggedright\footnotesize\vspace{4pt}
  \textsuperscript{a}Grounded in empathic accuracy~\cite{ickes1993empathic} and perspective mistaking~\cite{eyal2018mind}.
  \textsuperscript{b}Combines cognitive reflection from the RRQ with behavioral approach~\cite{elliot2006hierarchical}.
  \textsuperscript{c}Maps to the scaffolding functions described by Wood et al.~\cite{wood1976tutoring}.
\end{table}

\paragraph{Behavioral logs.}
\label{sec:logs}
The system recorded accept, modify, and reject actions for each beat. It also logged emotion-label changes, dwell time, and navigation events. We analyzed these data descriptively to identify interaction patterns.

\paragraph{Interview protocol.}
\label{sec:interview-protocol}

Semi-structured interviews lasted about 20 minutes per couple and addressed editing decisions, moments of surprise during Together Viewing, and differences from rereading chat logs. Interviews were audio-recorded and transcribed verbatim. Appendix~\ref{app:interview} provides the full protocol. 
We analyzed the transcripts using thematic analysis~\cite{braun2006using}. Two researchers independently coded all transcripts and developed themes through iterative comparison and discussion until reaching consensus~\cite{mcdonald2019reliability}. We mapped the themes to the research questions and examined behavioral logs for supporting and discrepant evidence.

\section{Findings}
\label{sec:findings}

\begin{figure}[t]
  \centering
  \includegraphics[width=\columnwidth]{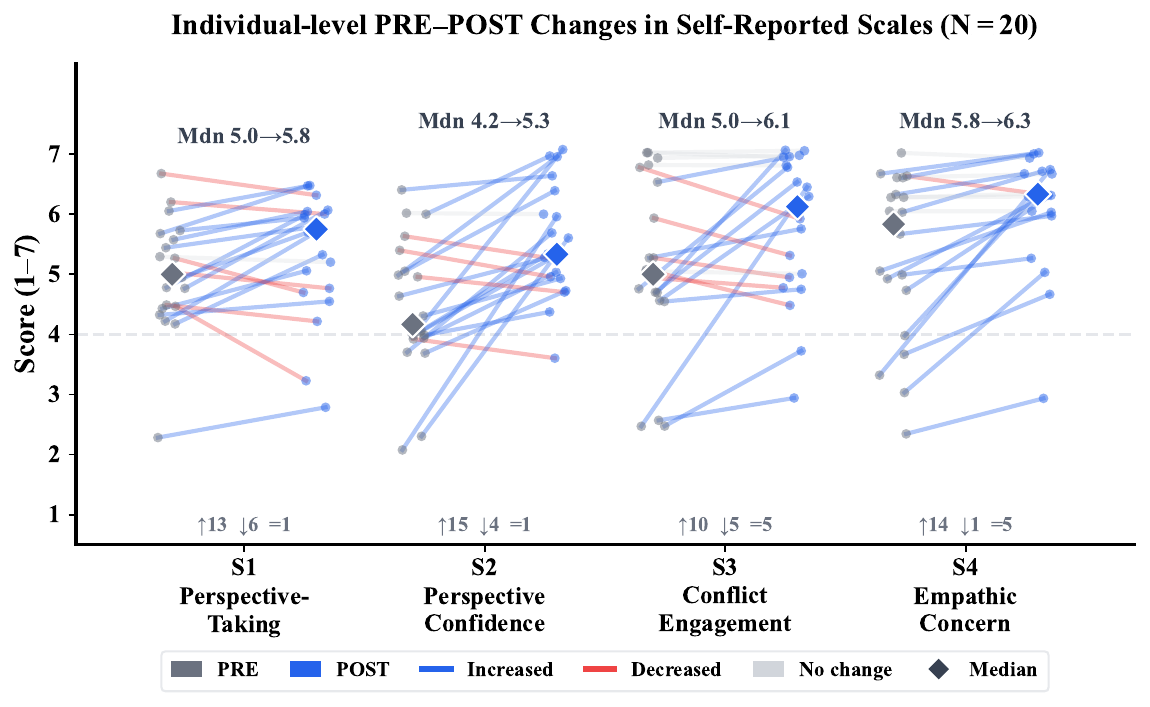}
  \caption{PRE-to-POST questionnaire results for S1--S4. Each line connects one participant's scores.}
  \Description{Paired dot plots show four scales with lines connecting each participant's PRE and POST scores.}
  \label{fig:scales}
\end{figure}

We organize the findings around the two research questions and the corresponding phases of DSR. Figure~\ref{fig:scales} summarizes the PRE-to-POST questionnaire results.
\revone{Because the PRE/POST scores reflect the complete ASIDE workflow, we interpret them as exploratory session-level patterns and make no phase-specific attribution.}

\revone{At the participant level, the median Perspective-Taking score (S1) changed from 5.00 to 5.75 ($t(19)=2.03$, $p=.056$, $d_z=0.45$), with increases for 13 participants. Perspective Confidence (S2) changed from 4.17 to 5.33 ($t(19)=3.93$, $p=.001$, $d_z=0.88$), with increases for 15 participants. Conflict Engagement (S3) changed from 5.00 to 6.12 ($W=26.5$, $p=.064$, $r_{\mathrm{rb}}=0.56$), with increases for 10 participants. Empathic Concern (S4) changed from 5.83 to 6.33 ($W=5.5$, $p=.001$, $r_{\mathrm{rb}}=0.91$), with increases for 14 participants.}
\revone{The dyad-level sensitivity analysis yielded results consistent with the participant-level analysis, with S1 ($t(9)=1.83$, $p=.101$, $d_z=0.58$), S2 ($t(9)=4.06$, $p=.003$, $d_z=1.28$), S3 ($t(9)=2.14$, $p=.061$, $d_z=0.68$), and S4 ($t(9)=3.23$, $p=.010$, $d_z=1.02$).}

The POST-only Scaffolding Effectiveness items (S5) further characterized participants' experience of the editing process. Participants rated the initial descriptions as helpful for starting reflection about their partner's feelings ($Mdn=6.0$), found editing them easier than writing from scratch ($Mdn=7.0$), and reported that even inaccurate descriptions could contribute to understanding ($Mdn=6.0$). The following sections draw on interviews and interaction logs to characterize participants' experiences during editing (RQ1) and joint reveal and viewing (RQ2).

\subsection{Reconstructing the Conflict from Within}
\label{sec:findings-rq1}

Participants made 262 edit decisions across all sessions, including 142 during Self-Confirm and 120 during Cross-Editing. Of the 142 Self-Confirm decisions, participants accepted 92 (64.8\%), modified 42 (29.6\%), and rejected 8 (5.6\%) AI-generated inferences. Of the 120 Cross-Editing decisions, they accepted 83 (69.2\%), modified 28 (23.3\%), and rejected 9 (7.5\%).

\subsubsection{AI Scaffolding Through Productive Inaccuracy}
\label{sec:findings-scaffold}

The editable AI drafts gave participants a concrete starting point for self-articulation.
P10a explained, ``If starting from zero, this matter would just get flipped past, but with this AI, you involuntarily want to analyze it more deeply.''
Several participants also found that the suggested emotional vocabulary helped them name feelings they had sensed but could not express (P3a, P5a, P10a).

Inaccurate inferences could also support articulation by giving users an interpretation to assess. P8a explained, ``People don't know what the right answer is, but they know very clearly what is wrong.'' P9a similarly found that an inaccurate description made the actual feeling immediately clearer. 
P7a described errors as generative. They made the process feel more targeted and elicited ``the natural urge to correct things.'' P7b added that this correction process produced ``a feeling of being guided toward reflection that reading chat logs can't give you.''
The scaffold's sentence-level structure also shaped what participants attended to. P4a noted that the system ``reviews sentence by sentence, making you think seriously about things you never considered at the time,'' and P4b, typically avoidant, found that the structure made it harder to disengage, saying, ``it's almost like it won't let you escape.''

The AI scaffold was not uniformly helpful.
P1b felt that the AI made the conflict sound more severe, explaining that ``the tone came out heavier than what we actually said.''
P8b found the inferences stayed at the surface:
``I expected the system to point to something more
core\ldots\ the deeper layers didn't come through.''
These cases illustrate a trade-off in the scaffold design. Recognizable inaccuracies can prompt correction, but they can also distort tone or remain shallow.

\subsubsection{Editing the Self: The Gap Between Words and Feelings}
\label{sec:findings-self}
The most consistent pattern in self-editing was the discovery of a gap between what participants had \emph{said} during the conflict and what they had actually \emph{felt}.
P8a distinguished the two layers sharply: ``The chat logs are the emotional byproduct of the moment---more like a weapon we used to attack each other during the conflict. What I wrote during editing, those inner monologues, that's what I
was really thinking. In this system, it might be real heart-to-heart communication.''
P9a's editing behavior illustrated the same gap. She revised 9 of her own inferences, explaining, ``I was really saying the opposite of what I meant.''
In one beat, she replaced the AI's grievance-focused interpretation with ``Hurry up and say you care about me. I'm going to run away from home!''
Her edits reframed statements that sounded angry as requests for reassurance. Similar edits often moved toward more vulnerable or honest expression rather than intensifying aggression.

Participants also described the indirect editing format as making self-disclosure easier. P5a noted that ``some feelings are too embarrassing to express directly, but this system is more indirect, so it's closer to your real inner self.''
Continued engagement with the editing process appeared to deepen this willingness. As P8a explained, ``There's no need to hide anything. I wanted to lay bare what I was truly thinking.''

\subsubsection{Editing the Partner: From Correction to Construction}
\label{sec:findings-partner-edit}
When participants edited the AI's inferences about their partner, the task shifted from checking a first-person account to constructing a plausible reading of another person's experience.
P6a contrasted the two tasks directly. She described self-editing as error correction, whereas engaging with the partner's perspective was where ``the real value'' lay.
The beat-by-beat reconstruction stripped away side conversations and surfaced the conflict's core logic. 
P6a said that it connected the conflict into ``one single thread'' and helped her understand how the partner arrived at each response.
By tracing this thread, participants shifted from knowing what the partner thought to understanding why.
P4a found that it helped her consider more objectively how the partner might see the conflict. P2b and P5b similarly described gaining ``a completely different angle'' and ``a new way of thinking about it''.

Participants also revised fault-focused interpretations into accounts that foregrounded the partner's vulnerability.
For example, P8a changed ``she doesn't actually feel bad, she's just afraid I'll leave'' to ``she's being too hard on herself, she's not that bad.'' This revision shifted the account from an attribution of bad intent toward protectiveness.

\subsection{Co-Viewing the Conflict from Outside}
\label{sec:findings-rq2}

\subsubsection{Moving from Participants to Co-Observers}
\label{sec:findings-frame}

\revoneR{Participants across nine pairs described the theatrical replay as changing their position relative to the conflict. Watching two figures enact their argument made a third-person stance more accessible than during direct chat-log rereading.}
P4a explained that ``the spectator mindset lifted us out of our emotions,'' allowing the partners to step outside their roles and view themselves as ``two little figures.''
P8a similarly described the experience as ``like watching a cartoon, completely third-person,'' which helped her think more rationally.

At the same time, the theatrical form gave the text a body and face, restoring a sense of embodiment that participants found missing from chat logs. P3b described ``a stronger sense of immersion'' because the scene presented ``a concrete person'' rather than flat text. P3a found that imagining a more concrete person made her ``slightly more restrained.'' P4b moved from finding the replay ``a bit funny, a bit ironic'' to feeling heartache during the same viewing. 

Participants therefore felt distanced from their roles while remaining immersed in the scene.
P7a explained that the figures ``bring us back into the event,'' while ``the observer aspect'' helped them think more clearly.
P3b also found that the contrast between cute characters and harsh words prompted reflection on his language. He explained, ``You wouldn't want such hurtful words to come from something so cute.'' 
\revoneR{Together, these accounts suggest that theatrical abstraction supported an observer stance while drawing participants back into the emotional content of the conflict.}

\subsubsection{Surfacing What the Conflict Had Hidden}
\label{sec:findings-revelation}

When both annotation sets became visible, participants encountered feelings that had remained hidden during the conflict. P4a explained that ``we tend to receive only the aggressive fragments,'' whereas the reveal showed ``the multiple facets of their emotions.''
The effects were specific.
P9a discovered that she had understood ``let's cool down for a bit'' as the start of an extended period of silence, whereas her partner intended only a brief pause to collect himself before continuing the conversation.
P5b had considered his partner's anger disproportionate, asking, ``Is it really worth getting so angry about?'' After seeing her account, he felt that her reaction was justified.
P10a found that a conflict the couple had considered resolved ten days earlier still contained a misunderstanding. She explained that ``through this system, he finally got the emotions we thought we'd already processed.''

\revtwoR{The synchronized reveal during Together Viewing was what P9a described as ``the most striking part'' of the experience. As both edited accounts came together in the shared view, participants' focus widened from individual beats and disputed phrases toward a fuller understanding of who their partner was beyond the immediate argument and how the conflict had unfolded within their relationship.
P10a explained, ``Through this system, I can understand my boyfriend more clearly. He really isn't the way I imagined when I was angry.'' P3b contrasted solo AI use, where he wanted the AI to ``judge who's right,'' with ASIDE, which felt ``more like being drawn into perspective-taking.'' His account reflects a shift from assigning blame to viewing conflict as a relational gap that both partners needed to bridge.}

\subsubsection{Continuing from Reveal to Conversation}
\label{sec:findings-conversation}

The reveal created an opening for further conversation. P9a explained, ``After hanging up from the experiment, we still needed to keep discussing.'' 
P1b similarly became curious about her partner's experience and explained, ``Once I started considering his perspective, I actually wanted to talk about what he was thinking at the time.'' \revtwoR{Divergence Cards gave this follow-up a beat-level reference by placing a self-confirmed inner state beside the partner's interpretation. This comparison provided a concrete starting point for discussing how their understandings had diverged.}

Participants linked this conversational opening to the credibility created by dual participation. P5b explained that the revealed content was ``based on content the other person has described and edited, so it's definitely closer to their real thoughts.'' P6a regarded participation itself as a signal of willingness. She observed, ``Two people participating means both are willing to face the problem; that attitude alone solves half of it.'' This shared commitment made the subsequent conversation easier to begin. P10a drew a similar contrast from her everyday use of the AI roleplay companion Hoshino for emotional support. The companion ``has no real emotions'' and knows only what she narrates to it, whereas ASIDE was grounded in ``the real thing that actually happened between us.''

Some participants also raised concerns about the effort and limits of the process. P2a felt that ``this small thing doesn't need this much time,'' while P3a found sentence-level annotation tedious. P6b experienced the process as an intrusion into the couple's private space and explained that ``it's a matter between the two of us, and suddenly it felt like a third party had entered.'' P9a questioned whether insight would necessarily change future behavior, noting that ``even though I know all the reasons, the next time it happens I'll still be the same way.'' These responses suggest that DSR may be more appropriate when the unresolved significance of a conflict justifies the effort involved.

ASIDE nevertheless prompted reflection beyond the immediate session. P3b said that he would reconsider future conflicts ``one more time'' before reacting. After recognizing how much emotional context could be lost in text, he planned to ``try to argue by voice, not text.'' The system did not resolve conflicts directly. Rather, the pixel-art staging entered participants' mental models and remained available as a frame for reinterpreting future emotional episodes.

\section{Discussion}
\label{sec:discussion}

\subsection{Productive Inaccuracy and Its Boundaries}
\label{sec:disc-productive-inaccuracy}

One source of ASIDE's value was the interaction structure through which partners revised and encountered AI-generated hypotheses. Relational AI systems often present AI as the source of insight through simulated interlocutors~\cite{shaikh2024rehearsal}, cognitive reframing~\cite{sharma-etal-2023-cognitive}, or emotional support based on one person's account.
ASIDE takes a different approach by presenting AI-generated inner states as provisional hypotheses that both partners can revise~\cite{wood1976tutoring,suh2023sensecape}.
Participants described these drafts as lowering the barrier to beginning reflection from a blank page (Section~\ref{sec:findings-scaffold}). Prior work on expressive writing further shows that articulating difficult experiences can support emotional processing~\cite{pennebaker1997writing}.
Authority over what either partner actually felt remains with the partners themselves. The output gains meaning through their participation~\cite{jiang2026}. This approach supports what Zhang et al.~\cite{10.1145/3772363.3798308} call relational sovereignty because the couple retains authorship over its emotional work.
Correcting recognizable errors was one route through which the editable scaffold supported articulation. Accuracy-oriented systems judge whether AI captures the user's experience~\cite{sharma2023human,nepal2024mindscape}. Our findings suggest a complementary question about whether an output activates the user's own knowledge. A precise inference may require little response, while a recognizable error can prompt the user to explain what they actually felt. Scaffolding theory locates value in this response~\cite{wood1976tutoring}.

Self-editing surfaced discrepancies between conflict language and felt experience. Articulating feelings can support affective processing~\cite{lieberman2007putting}, while mediated expression may lower barriers to disclosure~\cite{walther1996computer}. In ASIDE, participants used editable drafts to revise conflict-shaped language and articulate more vulnerable concerns.
This productive role differs across editing contexts. Users can evaluate self-inferences against their own experience and correct them directly. A partner's feelings cannot be retrieved in the same way, so participants must construct a reading from incomplete evidence~\cite{eyal2018mind,ickes1993empathic}. This makes partner inferences more vulnerable to anchoring~\cite{tversky1974judgment}.
\revtwo{Prior work often treats AI anchoring as a uniform risk~\cite{rastogi2022deciding,buccinca2021trust}. Reflective systems should calibrate uncertainty according to users' epistemic access to the inferred experience.}

\subsection{The Spectator Position}
\label{sec:disc-scaffolding}

\revoneR{Self-distancing research describes third-person perspective as a shift in the standpoint from which an experience is interpreted~\cite{kross2017self}. ASIDE extends this shift from individual reflection to a dyadic setting. Both partners are positioned outside the enacted conflict and in front of the same representation, creating a shared observer position. This arrangement may reduce the pressure to defend separate accounts and direct attention toward an event that both partners can inspect together.
This interpretation separates positional distance from representational vividness. Theatrical staging changes participants' relationship to their disputing roles, while embodied action preserves cues that support memory, emotional relevance, and engagement.} 
Non-photorealistic rendering can dampen emotional responses, although it may also reduce semantic clarity or interest ~\cite{mouldEmotionalResponseVisual2012}.

\revoneR{The scene and Divergence Cards supported complementary reflective actions. The scene embeds inner states in a continuous, embodied event, where timing, posture, and spatial staging support a shared observer position. The cards preserve specific interpretation gaps as beat-level comparisons that can support immediate discussion and later review.}
\revtwoR{Dyadic reflection systems can therefore adjust viewpoint and visual fidelity as separate design variables. Future comparisons could vary first- versus third-person staging and realistic versus stylized representation to examine their respective effects on defensiveness, recall, and willingness to discuss the conflict.}

\subsection{DSR as a Reusable Dyadic Structure}
\label{sec:disc-spectator}

\revtwo{Existing dyadic tools support joint engagement by structuring participation and exchange~\cite{ganong2012communication}.}
TogetherReflect~\cite{10.1145/3706598.3713642} supports separate emotional drawings followed by sharing. OurRelationship~\cite{doss2016} offers structured therapeutic modules, and PuppetChat~\cite{10.1145/3772318.3790685} structures dialogue through turn-taking. 
These systems make each partner's contribution visible, yet offer less support for inspecting how one partner interpreted the other at a particular moment. Mutual engagement alone does not make a specific misunderstanding locatable.

Dual participation in ASIDE contributed both content credibility and relational commitment. A partner's direct review made the revealed content more credible, while completing the process signaled willingness to face the conflict together. Prior work similarly illustrates the limits of individual participation for relationship-level outcomes~\cite{doss2016} and shows how lowering the threshold for reluctant partners can support engagement~\cite{cordova2014}. DSR reduces immediate interpersonal pressure through private, independently paced editing before the shared reveal. Its dependence on reciprocal participation is also a boundary. When only one partner is willing, the structure cannot provide mutual confirmation or a shared basis for reveal.

\revtwo{DSR brings these functions together as a reusable dyadic editing/reveal structure. Partners develop their accounts separately before encountering them together. Because both accounts refer to the same conversational beats, differences can be located within the original exchange. The reveal preserves the provenance of the AI-generated hypothesis, one partner's interpretation, and the other partner's confirmed account. DSR therefore offers a design structure that future systems can adapt to support shared reflection around interpersonal conflict.}

\subsection{Limitations and Future Work}
\label{sec:disc-limits}

\paragraph{\revtwo{Study scope and generalizability.}}
\revtwobody{The 10 voluntary young-adult couples provided paired perspectives, real conflict logs, interaction records, and interviews for an exploratory characterization of DSR. The small and relatively young sample limits generalization beyond this initial context. Future studies should examine broader age groups, non-romantic dyads, and different cultural and relationship settings.}

\paragraph{\revone{Whole-system attribution.}}
\revonebody{The PRE baseline followed participants' everyday processing of the conflict and an independent rereading of the chat log. Without a matched control, PRE-to-POST changes cannot separate ASIDE from the effects of structured reflection time. Future comparisons with guided journaling and individual AI reflection could test what reciprocal participation adds to structured individual reflection.}

\paragraph{\revoneR{Component mechanisms and representation.}}
\revtwoR{The interviews suggest different roles for theatrical replay and Divergence Cards, but controlled comparisons are needed to estimate their respective contributions. Future studies could hold the editing/reveal structure constant while comparing theatrical replay with a non-theatrical shared display. Holding the replay constant while adding or removing Divergence Cards could test their contribution to subsequent discussion and later review. The optimal abstraction level also remains unknown. First- and third-person viewpoints, realistic and stylized characters, and alternative staging could be varied independently to examine their effects on defensiveness, recall, and engagement.} 
\revoneR{Same-room staging and background selection may also shape interpretation when they differ from participants' remembered context.}

\paragraph{\revtwo{Safety requirements for dyadic AI reflection.}}
\revtwobody{This study excluded severe or unsafe conflicts because it did not provide clinical or crisis support. More broadly, systems that externalize and jointly reveal inferred inner states introduce risks across dyadic contexts. Power asymmetry may compromise voluntary participation, unresolved resentment may turn AI inferences into accusations, and unequal willingness may create asymmetric disclosure. Safety should therefore be addressed before and during use. Before participation, systems should assess each person's willingness and emotional readiness independently, and delay or decline use when reciprocal participation is uncertain. During interaction, partners should be able to pause, exit, slow down, and withhold specific confirmed states from disclosure. Systems should privately flag coercive or highly escalated input and suspend joint reveal when needed. 
AI-generated inner states should remain tentative hypotheses. Partner-confirmed accounts should be treated as authoritative regarding that partner's reported inner state, while remaining open to later revision.
Data governance should minimize collection, separate raw logs from reflective annotations, allow partners to delete their contributions, define retention periods, and prohibit secondary use without explicit consent. These safeguards should remain available throughout the reflective process.}

\paragraph{Persistence and transferability.}
A single session cannot show whether the observed changes persist or influence later behavior. Longitudinal studies should examine whether partners carry the observer position into later conflicts and return to recorded interpretation gaps. Repeated deployments should also assess changes in communication practices and possible reliance on system-mediated reflection. \revthree{The study was conducted in Chinese, including the LLM prompts. Because model outputs can reflect language-specific and cultural tendencies~\cite{lu2025cultural}, future studies should examine whether the prompting approach and findings transfer across languages and cultural settings.}
\section{Conclusion}
This paper introduced Dyadic Spectator Reflection (DSR), an interaction structure that repositions partners as co-observers of a past conflict, and ASIDE, a system that operationalizes DSR through editable AI hypotheses, independent editing, synchronized reveal, and theatrical replay. In an exploratory study with 10 couples, participants described the editable hypotheses as prompts for self-articulation, the joint reveal as making divergent readings visible, and the theatrical replay as helping them step out of their roles while remaining engaged with the event. DSR offers a reusable dyadic editing/reveal structure for future systems that support shared reflection around interpersonal conflict.
It positions AI as a revisable scaffold whose value emerges through partners' participation in examining and revising the interpretations it offers.

\section*{Generative AI Disclosure}
\revthree{This work involved Claude~Opus~4.6, Claude~Sonnet~4.6, and Gemini~2.5~Flash. The Claude models assisted with prototype development, debugging, and language refinement. Gemini~2.5~Flash transformed chat logs and participant-provided context into structured scenario specifications, including revisable inner-state hypotheses, context-matched background presets, and staging metadata. The authors reviewed and revised all AI-assisted outputs and take full responsibility for the system, analysis, and manuscript.}


\begin{acks}
We thank the participants for sharing their time, experiences, and perspectives. We also thank the anonymous reviewers and committee members for their thoughtful feedback, which helped strengthen this work.
\end{acks}


\bibliographystyle{ACM-Reference-Format}
\bibliography{reference}

@book{gottman1999seven,
  title={The Seven Principles for Making Marriage Work},
  author={Gottman, John M. and Silver, Nan},
  year={1999},
  publisher={Crown Publishers}
}

@book{hochschild1983managed,
  title={The Managed Heart: Commercialization of Human Feeling},
  author={Hochschild, Arlie Russell},
  year={1983},
  publisher={University of California Press}
}

@article{sillars2000cognition,
  title={Cognition during Marital Conflict: The Relationship of Thought and Talk},
  author={Sillars, Alan L. and Roberts, Linda J. and Leonard, Kenneth E. and Dun, Tim},
  journal={Journal of Social and Personal Relationships},
  volume={17},
  number={4--5},
  pages={479--502},
  year={2000},
  publisher={Sage}
}

@article{kenny2001accuracy,
  title={Accuracy and Bias in the Perception of the Partner in a Close Relationship},
  author={Kenny, David A. and Acitelli, Linda K.},
  journal={Journal of Personality and Social Psychology},
  volume={80},
  number={3},
  pages={439--448},
  year={2001},
  publisher={APA}
}

@article{bradbury1990attributions,
  title={Attributions in Marriage: Review and Critique},
  author={Bradbury, Thomas N. and Fincham, Frank D.},
  journal={Psychological Bulletin},
  volume={107},
  number={1},
  pages={3--33},
  year={1990},
  publisher={APA}
}

@article{fincham2004forgiveness,
  title={Forgiveness and Conflict Resolution in Marriage},
  author={Fincham, Frank D. and Beach, Steven R. H. and Davila, Joanne},
  journal={Journal of Family Psychology},
  volume={18},
  number={1},
  pages={72--81},
  year={2004},
  publisher={APA}
}

@article{eyal2018mind,
  title={Perspective Mistaking: Accurately Understanding the Mind of Another Requires Getting Perspective, Not Taking Perspective},
  author={Eyal, Tal and Steffel, Mary and Epley, Nicholas},
  journal={Journal of Personality and Social Psychology},
  volume={114},
  number={4},
  pages={547--571},
  year={2018},
  publisher={APA}
}

@article{epley2004perspective,
  title={Perspective Taking as Egocentric Anchoring and Adjustment},
  author={Epley, Nicholas and Keysar, Boaz and Van Boven, Leaf and Gilovich, Thomas},
  journal={Journal of Personality and Social Psychology},
  volume={87},
  number={3},
  pages={327--339},
  year={2004},
  publisher={APA}
}

@article{loewenstein2005hot,
  title={Hot--Cold Empathy Gaps and Medical Decision Making},
  author={Loewenstein, George},
  journal={Health Psychology},
  volume={24},
  number={4S},
  pages={S49--S56},
  year={2005},
  publisher={APA}
}

@article{lieberman2007putting,
  title={Putting Feelings into Words: Affect Labeling Disrupts Amygdala Activity in Response to Affective Stimuli},
  author={Lieberman, Matthew D. and Eisenberger, Naomi I. and Crockett, Molly J. and Tom, Sabrina M. and Pfeifer, Jennifer H. and Way, Baldwin M.},
  journal={Psychological Science},
  volume={18},
  number={5},
  pages={421--428},
  year={2007},
  publisher={Sage}
}

@article{ickes1993empathic,
  title={Empathic Accuracy},
  author={Ickes, William},
  journal={Journal of Personality},
  volume={61},
  number={4},
  pages={587--610},
  year={1993},
  publisher={Wiley}
}

@incollection{sillars1998misunderstanding,
  title={(Mis)understanding},
  author={Sillars, Alan L.},
  booktitle={The Dark Side of Close Relationships},
  editor={Spitzberg, Brian H. and Cupach, William R.},
  pages={73--102},
  year={1998},
  publisher={Lawrence Erlbaum Associates}
}

@incollection{culnan1987social,
  author    = {Culnan, Mary J. and Markus, M. Lynne},
  title     = {Information Technologies},
  booktitle = {Handbook of Organizational Communication: An Interdisciplinary Perspective},
  editor    = {Jablin, Fredric M. and Putnam, Linda L. and Roberts, Karlene H. and Porter, Lyman W.},
  pages     = {420--443},
  year      = {1987},
  publisher = {Sage Publications},
  address   = {Newbury Park, CA}
}

@article{walther1996computer,
  title={Computer-Mediated Communication: Impersonal, Interpersonal, and Hyperpersonal Interaction},
  author={Walther, Joseph B.},
  journal={Communication Research},
  volume={23},
  number={1},
  pages={3--43},
  year={1996},
  publisher={Sage}
}

@inproceedings{10.1145/2556288.2557177,
author = {Scissors, Lauren E. and Roloff, Michael E and Gergle, Darren},
title = {Room for interpretation: the role of self-esteem and CMC in romantic couple conflict},
year = {2014},
isbn = {9781450324731},
publisher = {Association for Computing Machinery},
address = {New York, NY, USA},
url = {https://doi.org/10.1145/2556288.2557177},
doi = {10.1145/2556288.2557177},
booktitle = {Proceedings of the SIGCHI Conference on Human Factors in Computing Systems},
pages = {3953–3962},
numpages = {10},
location = {Toronto, Ontario, Canada},
}

@article{noller1980misunderstandings,
  title={Misunderstandings in Marital Communication: A Study of Couples' Nonverbal Communication},
  author={Noller, Patricia},
  journal={Journal of Personality and Social Psychology},
  volume={39},
  number={6},
  pages={1135--1148},
  year={1980},
  publisher={APA}
}

@article{kross2011self,
  title={Making Meaning out of Negative Experiences by Self-Distancing},
  author={Kross, Ethan and Ayduk, {\"O}zlem},
  journal={Current Directions in Psychological Science},
  volume={20},
  number={3},
  pages={187--191},
  year={2011},
  publisher={Sage}
}

@incollection{kross2017self,
  title={Self-Distancing: Theory, Research, and Current Directions},
  author={Kross, Ethan and Ayduk, {\"O}zlem},
  booktitle={Advances in Experimental Social Psychology},
  volume={55},
  pages={81--136},
  year={2017},
  publisher={Academic Press}
}

@article{trope2010construal,
  title={Construal-Level Theory of Psychological Distance},
  author={Trope, Yaacov and Liberman, Nira},
  journal={Psychological Review},
  volume={117},
  number={2},
  pages={440--463},
  year={2010},
  publisher={APA}
}

@article{wood1976tutoring,
  title={The Role of Tutoring in Problem Solving},
  author={Wood, David and Bruner, Jerome S. and Ross, Gail},
  journal={Journal of Child Psychology and Psychiatry},
  volume={17},
  number={2},
  pages={89--100},
  year={1976},
  publisher={Wiley}
}

@article{tversky1974judgment,
  title={Judgment under Uncertainty: Heuristics and Biases},
  author={Tversky, Amos and Kahneman, Daniel},
  journal={Science},
  volume={185},
  number={4157},
  pages={1124--1131},
  year={1974},
  publisher={AAAS}
}

@book{johansen1988groupware,
  title={Groupware: Computer Support for Business Teams},
  author={Johansen, Robert},
  year={1988},
  publisher={Free Press}
}

@article{star1989institutional,
  title={Institutional Ecology, `Translations' and Boundary Objects: Amateurs and Professionals in Berkeley's Museum of Vertebrate Zoology, 1907--39},
  author={Star, Susan Leigh and Griesemer, James R.},
  journal={Social Studies of Science},
  volume={19},
  number={3},
  pages={387--420},
  year={1989},
  publisher={Sage}
}

@inproceedings{brereton2000observational,
  title={An observational study of how objects support engineering design thinking and communication: implications for the design of tangible media},
  author={Brereton, Margot and McGarry, Ben},
  booktitle={Proceedings of the SIGCHI conference on Human Factors in Computing Systems},
  pages={217--224},
  year={2000}
}

@book{schon1983reflective,
  title={The Reflective Practitioner: How Professionals Think in Action},
  author={Sch{\"o}n, Donald A.},
  year={1983},
  publisher={Basic Books}
}

@inproceedings{sengers2005reflective,
  title={Reflective Design},
  author={Sengers, Phoebe and Boehner, Kirsten and David, Shay and Kaye, Joseph `Jofish'},
  booktitle={Proceedings of the 4th Decennial Conference on Critical Computing},
  pages={49--58},
  year={2005},
  publisher={ACM}
}

@inproceedings{li2010stage,
  title={A Stage-Based Model of Personal Informatics Systems},
  author={Li, Ian and Dey, Anind and Forlizzi, Jodi},
  booktitle={Proceedings of the SIGCHI Conference on Human Factors in Computing Systems},
  pages={557--566},
  year={2010},
  publisher={ACM}
}

@inproceedings{rooksby2014lived,
  title={Personal Tracking as Lived Informatics},
  author={Rooksby, John and Rost, Mattias and Morrison, Alistair and Chalmers, Matthew},
  booktitle={Proceedings of the SIGCHI Conference on Human Factors in Computing Systems},
  pages={1163--1172},
  year={2014},
  publisher={ACM}
}

@inproceedings{epstein2015lived,
  title={A Lived Informatics Model of Personal Informatics},
  author={Epstein, Daniel A. and Ping, An and Fogarty, James and Munson, Sean A.},
  booktitle={Proceedings of the 2015 ACM International Joint Conference on Pervasive and Ubiquitous Computing},
  pages={731--742},
  year={2015},
  publisher={ACM}
}

@inproceedings{kim2024mindfuldiary,
  title={MindfulDiary: Harnessing Large Language Model to Support Psychiatric Patients' Journaling},
  author={Kim, Taewan and Bae, Seolyeong and Kim, Hyun Ah and Lee, Su-Woo and Hong, Hwajung and Yang, Chanmo and Kim, Young-Ho},
  booktitle={Proceedings of the CHI Conference on Human Factors in Computing Systems},
  year={2024},
  publisher={ACM},
  doi={10.1145/3613904.3642937}
}

@inproceedings{10.1145/3613904.3642693,
author = {Kim, Taewan and Shin, Donghoon and Kim, Young-Ho and Hong, Hwajung},
title = {DiaryMate: Understanding User Perceptions and Experience in Human-AI Collaboration for Personal Journaling},
year = {2024},
isbn = {9798400703300},
publisher = {Association for Computing Machinery},
address = {New York, NY, USA},
url = {https://doi.org/10.1145/3613904.3642693},
doi = {10.1145/3613904.3642693},
booktitle = {Proceedings of the 2024 CHI Conference on Human Factors in Computing Systems},
articleno = {1046},
numpages = {15},
location = {Honolulu, HI, USA},
}

@inproceedings{nepal2024mindscape,
  title={Contextual AI Journaling: Integrating LLM and Time Series Behavioral Sensing Technology to Promote Self-Reflection and Well-being using the MindScape App},
  author={Nepal, Subigya and Pillai, Arvind and Campbell, William and Massachi, Talie and Choi, Eunsol Soul and Xu, Orson and Kuc, Joanna and Huckins, Jeremy and Holden, Jason and Depp, Colin and Jacobson, Nicholas and Czerwinski, Mary and Granholm, Eric and Campbell, Andrew T.},
  booktitle={Extended Abstracts of the CHI Conference on Human Factors in Computing Systems},
  year={2024},
  publisher={ACM},
  doi={10.1145/3613905.3650767}
}

@inproceedings{isaacs2013echoes,
  title={Echoes from the Past: How Technology Mediated Reflection Improves Well-Being},
  author={Isaacs, Ellen and Konrad, Artie and Walendowski, Alan and Lennig, Thomas and Hollis, Victoria and Whittaker, Steve},
  booktitle={Proceedings of the SIGCHI Conference on Human Factors in Computing Systems},
  pages={1071--1080},
  year={2013},
  publisher={ACM}
}

@inproceedings{10.1145/3491102.3502030,
author = {Lee, Mina and Liang, Percy and Yang, Qian},
title = {CoAuthor: Designing a Human-AI Collaborative Writing Dataset for Exploring Language Model Capabilities},
year = {2022},
isbn = {9781450391573},
publisher = {Association for Computing Machinery},
address = {New York, NY, USA},
url = {https://doi.org/10.1145/3491102.3502030},
doi = {10.1145/3491102.3502030},
booktitle = {Proceedings of the 2022 CHI Conference on Human Factors in Computing Systems},
articleno = {388},
numpages = {19},
location = {New Orleans, LA, USA},
}

@inproceedings{gero2023social,
  author    = {Gero, Katy Ilonka and Long, Tao and Chilton, Lydia B.},
  title     = {Social Dynamics of {AI} Support in Creative Writing},
  booktitle = {Proceedings of the 2023 CHI Conference on Human Factors in Computing Systems},
  year      = {2023},
  articleno = {245},
  numpages  = {15},
  pages     = {1--15},
  publisher = {Association for Computing Machinery},
  address   = {New York, NY, USA},
  doi       = {10.1145/3544548.3580782}
}

@inproceedings{yuan2022wordcraft,
  title={Wordcraft: Story Writing with Large Language Models},
  author={Yuan, Ann and Coenen, Andy and Reif, Emily and Ippolito, Daphne},
  booktitle={Proceedings of the 27th International Conference on Intelligent User Interfaces},
  pages={841--852},
  year={2022},
  publisher={ACM}
}

@inproceedings{chung2022talebrush,
  author    = {Chung, John Joon Young and Kim, Wooseok and Yoo, Kang Min and Lee, Hwaran and Adar, Eytan and Chang, Minsuk},
  title     = {TaleBrush: Sketching Stories with Generative Pretrained Language Models},
  booktitle = {Proceedings of the 2022 CHI Conference on Human Factors in Computing Systems},
  year      = {2022},
  pages     = {1--19},
  publisher = {Association for Computing Machinery},
  address   = {New York, NY, USA},
  doi       = {10.1145/3491102.3501819}
}

@inproceedings{10.1145/3772318.3790572,
author = {Xu, Jiangnan and Cha, Haeseul and Choi, Gosu and Lee, Gyu-cheol and Yoon, Yeo-Jin and Lee, Zucheul and Papangelis, Konstantinos and Kim, Dae Hyun and Kim, Juho},
title = {DiaryPlay: AI-Assisted Creation of Interactive Story Vignettes for Everyday Storytelling},
year = {2026},
isbn = {9798400722783},
publisher = {Association for Computing Machinery},
address = {New York, NY, USA},
url = {https://doi.org/10.1145/3772318.3790572},
doi = {10.1145/3772318.3790572},
booktitle = {Proceedings of the 2026 CHI Conference on Human Factors in Computing Systems},
articleno = {924},
numpages = {21},
location = {
},
}

@article{fitzpatrick2017delivering,
  author  = {Fitzpatrick, Kathleen Kara and Darcy, Alison and Vierhile, Molly},
  title   = {Delivering Cognitive Behavior Therapy to Young Adults With Symptoms of Depression and Anxiety Using a Fully Automated Conversational Agent ({Woebot}): A Randomized Controlled Trial},
  journal = {JMIR Mental Health},
  year    = {2017},
  volume  = {4},
  number  = {2},
  pages   = {e19},
  doi     = {10.2196/mental.7785}
}

@article{skjuve2021my,
  title={My Chatbot Companion -- A Study of Human--Chatbot Relationships},
  author={Skjuve, Marita and F{\o}lstad, Asbjørn and Haugstveit, Ida Maria and Brandtzæg, Petter Bae},
  journal={International Journal of Human-Computer Studies},
  volume={149},
  pages={102601},
  year={2021},
  publisher={Elsevier}
}

@article{sharma2023human,
  title={Human--{AI} Collaboration Enables More Empathic Conversations in Text-Based Peer-to-Peer Mental Health Support},
  author={Sharma, Ashish and Lin, Inna W. and Miner, Adam S. and Atkins, David C. and Althoff, Tim},
  journal={Nature Machine Intelligence},
  volume={5},
  number={1},
  pages={46--57},
  year={2023},
  publisher={Nature Publishing Group}
}

@inproceedings{shaikh2024rehearsal,
  title={Rehearsal: Simulating Conflict to Teach Conflict Resolution},
  author={Shaikh, Omar and Chai, Valentino Emil and Gelfand, Michele and Yang, Diyi and Bernstein, Michael S.},
  booktitle={Proceedings of the CHI Conference on Human Factors in Computing Systems},
  year={2024},
  publisher={ACM},
  doi={10.1145/3613904.3642159}
}

@inproceedings{chun2025conflictlens,
  author    = {Chun, Jiwon and Zhang, Gefei and Xia, Meng},
  title     = {ConflictLens: LLM-Based Conflict Resolution Training in Romantic Relationship},
  booktitle = {Adjunct Proceedings of the 38th Annual ACM Symposium on User Interface Software and Technology},
  year      = {2025},
  pages     = {1--3},
  publisher = {Association for Computing Machinery},
  address   = {New York, NY, USA},
  doi       = {10.1145/3746058.3758422}
}

@inproceedings{park2023generative,
  title={Generative Agents: Interactive Simulacra of Human Behavior},
  author={Park, Joon Sung and O'Brien, Joseph C. and Cai, Carrie J. and Morris, Meredith Ringel and Liang, Percy and Bernstein, Michael S.},
  booktitle={Proceedings of the 36th Annual ACM Symposium on User Interface Software and Technology},
  year={2023},
  publisher={ACM},
  doi={10.1145/3586183.3606763}
}

@article{baughan2024hardconversations,
  title={Supporting Hard Conversations in Close Relationships Through Design},
  author={Baughan, Amanda and Tian, Larry and Shekar, Pranav and Zhang, Amy and Hiniker, Alexis},
  journal={Proceedings of the ACM on Human-Computer Interaction},
  volume={8},
  number={CSCW2},
  articleno={379},
  year={2024},
  publisher={ACM},
  doi={10.1145/3686918}
}

@inproceedings{10.1145/3706598.3713642,
author = {Wagener, Nadine and Albensoeder, Daniel Christian and Reicherts, Leon and Wo\'{z}niak, Pawe\l{} W. and Rogers, Yvonne and Niess, Jasmin},
title = {TogetherReflect: Supporting Emotional Expression in Couples Through a Collaborative Virtual Reality Experience},
year = {2025},
isbn = {9798400713941},
publisher = {Association for Computing Machinery},
address = {New York, NY, USA},
url = {https://doi.org/10.1145/3706598.3713642},
doi = {10.1145/3706598.3713642},
booktitle = {Proceedings of the 2025 CHI Conference on Human Factors in Computing Systems},
articleno = {1130},
numpages = {16},
location = {
},
}

@article{hancock2020aimc,
  title={{AI}-Mediated Communication: Definition, Research Agenda, and Ethical Considerations},
  author={Hancock, Jeffrey T. and Naaman, Mor and Levy, Karen},
  journal={Journal of Computer-Mediated Communication},
  volume={25},
  number={1},
  pages={89--100},
  year={2020},
  publisher={Oxford University Press}
}

@incollection{ebner2012games,
  title={Using Games to Teach Negotiation and Dispute Resolution},
  author={Ebner, Noam and Kovach, Kimberlee K.},
  booktitle={Assessing Our Students, Assessing Ourselves},
  editor={Honeyman, Christopher and Coben, James and De Palo, Giuseppe},
  year={2012},
  publisher={DRI Press}
}

@inproceedings{suh2023sensecape,
  title={Sensecape: Enabling Multilevel Exploration and Sensemaking with Large Language Models},
  author={Suh, Sangho and Min, Bryan and Palani, Srishti and Xia, Haijun},
  booktitle={The 36th Annual ACM Symposium on User Interface Software and Technology},
  year={2023},
  publisher={ACM},
  doi={10.1145/3586183.3606756}
}

@inproceedings{jiang2023graphologue,
  title={Graphologue: Exploring Large Language Model Responses with Interactive Diagrams},
  author={Jiang, Peiling and Rayan, Jude and Dow, Steven P. and Xia, Haijun},
  booktitle={The 36th Annual ACM Symposium on User Interface Software and Technology},
  year={2023},
  publisher={ACM},
  doi={10.1145/3586183.3606737}
}

@inproceedings{suh2024luminate,
  title={Luminate: Structured Generation and Exploration of Design Space with Large Language Models for Human-{AI} Co-Creation},
  author={Suh, Sangho and Chen, Meng and Min, Bryan and Li, Toby Jia-Jun and Xia, Haijun},
  booktitle={Proceedings of the CHI Conference on Human Factors in Computing Systems},
  year={2024},
  publisher={ACM},
  doi={10.1145/3613904.3642400}
}

@article{kruger2005email,
  title={Egocentrism over E-Mail: Can We Communicate as Well as We Think?},
  author={Kruger, Justin and Epley, Nicholas and Parker, Jason and Ng, Zhi-Wen},
  journal={Journal of Personality and Social Psychology},
  volume={89},
  number={6},
  pages={925--936},
  year={2005},
  publisher={American Psychological Association},
  doi={10.1037/0022-3514.89.6.925}
}

@article{ayduk2010distance,
  title={From a Distance: Implications of Spontaneous Self-Distancing for Adaptive Self-Reflection},
  author={Ayduk, {\"O}zlem and Kross, Ethan},
  journal={Journal of Personality and Social Psychology},
  volume={98},
  number={5},
  pages={809--829},
  year={2010},
  doi={10.1037/a0019205}
}

@article{grossmann2014solomon,
  title={Exploring Solomon's Paradox: Self-Distancing Eliminates the Self-Other Asymmetry in Wise Reasoning About Close Relationships in Younger and Older Adults},
  author={Grossmann, Igor and Kross, Ethan},
  journal={Psychological Science},
  volume={25},
  number={8},
  pages={1571--1580},
  year={2014},
  doi={10.1177/0956797614535400}
}

@article{ganong2012communication,
  title={Communication technology and postdivorce coparenting},
  author={Ganong, Lawrence H and Coleman, Marilyn and Feistman, Richard and Jamison, Tyler and Stafford Markham, Melinda},
  journal={Family relations},
  volume={61},
  number={3},
  pages={397--409},
  year={2012},
  publisher={Wiley Online Library}
}

@book{gottman2015principia,
  title={Principia Amoris: The New Science of Love},
  author={Gottman, John M.},
  year={2015},
  publisher={Routledge}
}

@article{suthers2001towards,
  author  = {Suthers, Daniel D.},
  title   = {Towards a Systematic Study of Representational Guidance for Collaborative Learning Discourse},
  journal = {Journal of Universal Computer Science},
  year    = {2001},
  volume  = {7},
  number  = {3},
  pages   = {254--277},
  doi     = {10.3217/jucs-007-03-0254}
}

@inproceedings{sharma2024towards,
  author    = {Sharma, Mrinank and Tong, Meg and Korbak, Tomasz and Duvenaud, David and Askell, Amanda and Bowman, Samuel R. and Cheng, Newton and Durmus, Esin and Hatfield-Dodds, Zac and Johnston, Scott R. and Kravec, Shauna M. and Maxwell, Timothy and McCandlish, Sam and Ndousse, Kamal and Rausch, Oliver and Schiefer, Nicholas and Yan, Da and Zhang, Miranda and Perez, Ethan},
  title     = {Towards Understanding Sycophancy in Language Models},
  booktitle = {International Conference on Learning Representations},
  year      = {2024},
  url       = {https://openreview.net/forum?id=tvhaxkMKAn}
}

@inproceedings{cuadra2024illusion,
  title={The Illusion of Empathy? Notes on Displays of Emotion in Human-Computer Interaction},
  author={Cuadra, Andrea and Wang, Maria and Stein, Lynn Andrea and Jung, Malte F. and Dell, Nicola and Estrin, Deborah and Landay, James A.},
  booktitle={Proceedings of the CHI Conference on Human Factors in Computing Systems},
  year={2024},
  publisher={ACM},
  doi={10.1145/3613904.3642336}
}

@article{ross1991reactive,
  title={Barriers to Conflict Resolution},
  author={Ross, Lee and Stillinger, Constance},
  journal={Negotiation Journal},
  volume={7},
  number={4},
  pages={389--404},
  year={1991},
  publisher={Wiley}
}

@article{nisbett1977telling,
  title={Telling More Than We Can Know: Verbal Reports on Mental Processes},
  author={Nisbett, Richard E. and Wilson, Timothy DeCamp},
  journal={Psychological Review},
  volume={84},
  number={3},
  pages={231--259},
  year={1977},
  publisher={APA}
}

@article{kross2014self,
  title={Self-Talk as a Regulatory Mechanism: How You Do It Matters},
  author={Kross, Ethan and Bruehlman-Senecal, Emma and Park, Jiyoung and Burson, Aleah and Dougherty, Adrienne and Shablack, Holly and Bremner, Ryan and Moser, Jason and Ayduk, {\"O}zlem},
  journal={Journal of Personality and Social Psychology},
  volume={106},
  number={2},
  pages={304--324},
  year={2014},
  publisher={APA}
}

@incollection{clark1991grounding,
  title={Grounding in Communication},
  author={Clark, Herbert H. and Brennan, Susan E.},
  booktitle={Perspectives on Socially Shared Cognition},
  editor={Resnick, Lauren B. and Levine, John M. and Teasley, Stephanie D.},
  pages={127--149},
  year={1991},
  publisher={American Psychological Association}
}

@book{mccloud1993understanding,
  title={Understanding Comics: The Invisible Art},
  author={McCloud, Scott},
  year={1993},
  publisher={William Morrow}
}

@book{hall1966hidden,
  title={The Hidden Dimension},
  author={Hall, Edward T.},
  year={1966},
  publisher={Doubleday}
}

@book{creswell2017research,
  author    = {Creswell, John W. and Creswell, J. David},
  title     = {Research Design: Qualitative, Quantitative, and Mixed Methods Approaches},
  edition   = {5},
  year      = {2018},
  publisher = {SAGE Publications},
  address   = {Thousand Oaks, CA},
  isbn      = {9781506386706}
}

@article{braun2006using,
  title={Using Thematic Analysis in Psychology},
  author={Braun, Virginia and Clarke, Victoria},
  journal={Qualitative Research in Psychology},
  volume={3},
  number={2},
  pages={77--101},
  year={2006},
  publisher={Taylor \& Francis},
  doi={10.1191/1478088706qp063oa}
}

@article{davis1983,
  title={Measuring Individual Differences in Empathy: Evidence for a Multidimensional Approach},
  author={Davis, Mark H.},
  journal={Journal of Personality and Social Psychology},
  volume={44},
  number={1},
  pages={113--126},
  year={1983},
  publisher={APA},
  doi={10.1037/0022-3514.44.1.113}
}

@article{trapnell1999,
  title={Private Self-Consciousness and the Five-Factor Model of Personality: Distinguishing Rumination from Reflection},
  author={Trapnell, Paul D. and Campbell, Jennifer D.},
  journal={Journal of Personality and Social Psychology},
  volume={76},
  number={2},
  pages={284--304},
  year={1999},
  publisher={APA},
  doi={10.1037/0022-3514.76.2.284}
}

@article{pennebaker1997writing,
  title={Writing about Emotional Experiences as a Therapeutic Process},
  author={Pennebaker, James W.},
  journal={Psychological Science},
  volume={8},
  number={3},
  pages={162--166},
  year={1997},
  publisher={Sage}
}

@inproceedings{sharma-etal-2023-cognitive,
    title = "Cognitive Reframing of Negative Thoughts through Human-Language Model Interaction",
    author = "Sharma, Ashish  and
      Rushton, Kevin  and
      Lin, Inna  and
      Wadden, David  and
      Lucas, Khendra  and
      Miner, Adam  and
      Nguyen, Theresa  and
      Althoff, Tim",
    editor = "Rogers, Anna  and
      Boyd-Graber, Jordan  and
      Okazaki, Naoaki",
    booktitle = "Proceedings of the 61st Annual Meeting of the Association for Computational Linguistics (Volume 1: Long Papers)",
    month = jul,
    year = "2023",
    address = "Toronto, Canada",
    publisher = "Association for Computational Linguistics",
    url = "https://aclanthology.org/2023.acl-long.555/",
    doi = "10.18653/v1/2023.acl-long.555",
    pages = "9977--10000",
}

@article{parkSteppingBackMove2016,
  author  = {Park, Jiyoung and Ayduk, {\"O}zlem and Kross, Ethan},
  title   = {Stepping Back to Move Forward: Expressive Writing Promotes Self-Distancing},
  journal = {Emotion},
  year    = {2016},
  volume  = {16},
  number  = {3},
  pages   = {349--364},
  doi     = {10.1037/emo0000121}
}

@article{christensen1990,
  title={Gender and Social Structure in the Demand/Withdraw Pattern of Marital Conflict},
  author={Christensen, Andrew and Heavey, Christopher L.},
  journal={Journal of Personality and Social Psychology},
  volume={59},
  number={1},
  pages={73--81},
  year={1990},
  publisher={APA},
  doi={10.1037/0022-3514.59.1.73}
}

@article{mouldEmotionalResponseVisual2012,
  author  = {Mould, David and Mandryk, Regan L. and Li, Hua},
  title   = {Emotional Response and Visual Attention to Non-Photorealistic Images},
  journal = {Computers \& Graphics},
  year    = {2012},
  volume  = {36},
  number  = {6},
  pages   = {658--672},
  doi     = {10.1016/j.cag.2012.03.039}
}

@inproceedings{jiang2026,
  author    = {Jiang, Zhuoqun and Yeo, ShunYi and Herremans, Dorien and Perrault, Simon Tangi},
  title     = {Scaffolded Vulnerability: Chatbot-Mediated Reciprocal Self-Disclosure and Need-Supportive Interaction in Couples},
  booktitle = {Proceedings of the 2026 CHI Conference on Human Factors in Computing Systems},
  year      = {2026},
  articleno = {1296},
  numpages  = {39},
  pages     = {1--39},
  publisher = {Association for Computing Machinery},
  address   = {New York, NY, USA},
  doi       = {10.1145/3772318.3791370}
}

@article{murray2002kindred,
  title={Kindred Spirits? The Benefits of Egocentrism in Close Relationships},
  author={Murray, Sandra L. and Holmes, John G. and Bellavia, Gina and Griffin, Dale W. and Dolderman, Dan},
  journal={Journal of Personality and Social Psychology},
  volume={82},
  number={4},
  pages={563--581},
  year={2002},
  publisher={APA},
  doi={10.1037/0022-3514.82.4.563}
}

@article{byron2008carrying,
  title={Carrying Too Heavy a Load? The Communication and Miscommunication of Emotion by Email},
  author={Byron, Kristin},
  journal={Academy of Management Review},
  volume={33},
  number={2},
  pages={309--327},
  year={2008},
  publisher={Academy of Management}
}

@book{johnson2004practice,
  title={The Practice of Emotionally Focused Couple Therapy: Creating Connection},
  author={Johnson, Susan M.},
  edition={2nd},
  year={2004},
  publisher={Brunner-Routledge}
}

@article{karney2000attributions,
  title={Attributions in Marriage: State or Trait? A Growth Curve Analysis},
  author={Karney, Benjamin R. and Bradbury, Thomas N.},
  journal={Journal of Personality and Social Psychology},
  volume={78},
  number={2},
  pages={295--309},
  year={2000},
  publisher={APA},
  doi={10.1037/0022-3514.78.2.295}
}

@book{ickes2003everyday,
  title={Everyday Mind Reading: Understanding What Other People Think and Feel},
  author={Ickes, William},
  year={2003},
  publisher={Prometheus Books}
}

@book{clark1996using,
  title={Using Language},
  author={Clark, Herbert H.},
  year={1996},
  publisher={Cambridge University Press}
}

@article{nils2012beyond,
  title={Beyond the myth of venting: Social sharing modes determine the benefits of emotional disclosure},
  author={Nils, Fr{\'e}d{\'e}ric and Rim{\'e}, Bernard},
  journal={European Journal of Social Psychology},
  volume={42},
  number={6},
  pages={672--681},
  year={2012},
  publisher={Wiley Online Library}
}

@inproceedings{10.1145/3772363.3798308,
author = {Zhang, Xinyi and Zhu, Zicheng and Su, Yuxin},
title = {"Clumsiness is Love Itself": How Long-Distance Couples Experience, Envision, and Resist {AI} in Intimate Relationships},
year = {2026},
isbn = {9798400722813},
publisher = {Association for Computing Machinery},
address = {New York, NY, USA},
url = {https://doi.org/10.1145/3772363.3798308},
doi = {10.1145/3772363.3798308},
booktitle = {Proceedings of the Extended Abstracts of the 2026 CHI Conference on Human Factors in Computing Systems},
articleno = {4},
pages = {1--7},
numpages = {7},
location = {
},
series = {CHI EA '26}
}

@article{rastogi2022deciding,
  title={Deciding fast and slow: The role of cognitive biases in ai-assisted decision-making},
  author={Rastogi, Charvi and Zhang, Yunfeng and Wei, Dennis and Varshney, Kush R and Dhurandhar, Amit and Tomsett, Richard},
  journal={Proceedings of the ACM on Human-computer Interaction},
  volume={6},
  number={CSCW1},
  pages={1--22},
  year={2022},
  publisher={ACM New York, NY, USA}
}

@article{buccinca2021trust,
  title={To trust or to think: cognitive forcing functions can reduce overreliance on AI in AI-assisted decision-making},
  author={Bu{\c{c}}inca, Zana and Malaya, Maja Barbara and Gajos, Krzysztof Z},
  journal={Proceedings of the ACM on Human-computer Interaction},
  volume={5},
  number={CSCW1},
  pages={1--21},
  year={2021},
  publisher={ACM New York, NY, USA}
}

@article{doss2016,
  author  = {Doss, Brian D. and Cicila, Larisa N. and Georgia, Emily J. and Roddy, McKenzie K. and Nowlan, Kathryn M. and Benson, Lisa A. and Christensen, Andrew},
  title   = {A Randomized Controlled Trial of the Web-Based {OurRelationship} Program: Effects on Relationship and Individual Functioning},
  journal = {Journal of Consulting and Clinical Psychology},
  year    = {2016},
  volume  = {84},
  number  = {4},
  pages   = {285--296},
  doi     = {10.1037/ccp0000063}
}

@article{cordova2014,
  author  = {Cordova, James V. and Eubanks Fleming, C. J. and Morrill, Melinda Ippolito and Hawrilenko, Matt and Sollenberger, Julia W. and Harp, Amanda G. and Gray, Tatiana D. and Darling, Ellen V. and Blair, Jonathan M. and Meade, Amy E. and Wachs, Karen},
  title   = {The Marriage Checkup: A Randomized Controlled Trial of Annual Relationship Health Checkups},
  journal = {Journal of Consulting and Clinical Psychology},
  year    = {2014},
  volume  = {82},
  number  = {4},
  pages   = {592--604},
  doi     = {10.1037/a0037097}
}

@inproceedings{10.1145/3772318.3790685,
author = {Wang, Emma Jiren and Hu, Siying and Lu, Zhicong},
title = {PuppetChat: Fostering Intimate Communication through Bidirectional Actions and Micronarratives},
year = {2026},
isbn = {9798400722783},
publisher = {Association for Computing Machinery},
address = {New York, NY, USA},
url = {https://doi.org/10.1145/3772318.3790685},
doi = {10.1145/3772318.3790685},
booktitle = {Proceedings of the 2026 CHI Conference on Human Factors in Computing Systems},
articleno = {1295},
numpages = {19},
location = {
},
}

@article{elliot2006hierarchical,
  title={The hierarchical model of approach-avoidance motivation},
  author={Elliot, Andrew J},
  journal={Motivation and emotion},
  volume={30},
  number={2},
  pages={111--116},
  year={2006},
  publisher={Springer}
}

@article{mcdonald2019reliability,
  title={Reliability and inter-rater reliability in qualitative research: Norms and guidelines for CSCW and HCI practice},
  author={McDonald, Nora and Schoenebeck, Sarita and Forte, Andrea},
  journal={Proceedings of the ACM on human-computer interaction},
  volume={3},
  number={CSCW},
  pages={1--23},
  year={2019},
  publisher={ACM New York, NY, USA}
}

@article{leonard2025navigating,
  author  = {L{\'e}onard, Florence and M{\'e}tellus, Sarafina and Vaillancourt-Morel, Marie-Pier and Brassard, Audrey and Margolin, Gayla and Daspe, Marie-{\`E}ve},
  title   = {Navigating Cyber Intimate Partner Violence and Conflict: Negative Anticipation and Emotions During Text-Based Versus Face-to-Face Conflict Discussions in Young Adult Couples},
  journal = {Cyberpsychology, Behavior, and Social Networking},
  year    = {2025},
  volume  = {28},
  number  = {5},
  pages   = {359--365},
  doi     = {10.1089/cyber.2024.0460}
}

@article{lu2025cultural,
  author = {Lu, Jackson G. and Song, Lesley Luyang and Zhang, Lu Doris},
  title = {Cultural Tendencies in Generative {AI}},
  journal = {Nature Human Behaviour},
  volume = {9},
  pages = {2360--2369},
  year = {2025},
  doi = {10.1038/s41562-025-02242-1}
}

\appendix
\section{Relationship Scripting Language (RSL)}
\label{app:rsl}

\revthree{Each conflict scenario is encoded in a structured schema called the Relationship Scripting Language (RSL). RSL separates \emph{immutable facts} about what was said from \emph{editable interpretations} of what may have been thought, while embedding both in a theatrical representation.}

\subsection{Schema}

A scenario document contains metadata, persona definitions, and an ordered sequence of beats:

\begin{Verbatim}[fontsize=\small, frame=single, commandchars=\\\{\}]
scenario := \{
  id:       string,
  title:    string,
  scene:    string,          \textrm{// scene preset key}
  personas: \{A: Persona, B: Persona\},
  beats:    [beat_0, ..., beat_n]
\}

beat := \{
  id:        int,
  intensity: float [0,1],    \textrm{// conflict arc position}
  narrator:  string,         \textrm{// factual stage direction}
  dialogue:  \{               \textrm{// verbatim chat text}
    speaker: 'A' | 'B',
    text:    string
  \},
  thoughts:  \{               \textrm{// AI-generated, user-editable}
    A: \{text, emotion\},
    B: \{text, emotion\}
  \},
  spatial:   \{               \textrm{// theatrical staging}
    A: \{x, facing, pose, scale\},
    B: \{x, facing, pose, scale\}
  \},
  proxemic:  \{state, divider\}  \textrm{// relational distance}
\}
\end{Verbatim}

\subsection{Field Specifications}

\paragraph{Fact layer (immutable).}
The \texttt{dialogue.text} field is copied verbatim from the original chat record and cannot be altered by the AI, the system, or either user. \revthree{Each beat includes a concise \texttt{narrator} line that describes timing, conversational progression, or observable actions without inferring either partner's emotion.}

\paragraph{Interpretation layer (editable).}
The \texttt{thoughts} field is AI-generated and user-editable through the accept, modify, and reject controls described in Section~\ref{sec:dyadic}. Each thought contains:
\begin{itemize}
  \item \texttt{text}: inner-state text in one to three first-person sentences; and
  \item \revthree{\texttt{emotion}: an emotion tag used to select the corresponding character pose and icon cue.}
\end{itemize}

\paragraph{Staging layer.}
The \texttt{spatial} field controls each character's position (\texttt{x}: 0--100\% horizontal), facing direction, body pose, and scale. The \texttt{proxemic} field records the relational distance state of the beat and whether a divider is drawn between the characters, following Hall's account of proxemic distance~\cite{hall1966hidden}. The \texttt{intensity} field represents the beat's position in the modeled conflict arc on a normalized 0--1 scale. It controls vignette strength, scene darkening, and the color temperature of the ambient particle layer.

\subsection{Design Rationale}

Table~\ref{tab:rsl-rationale} summarizes the design decisions encoded in each RSL field and their grounding.

\begin{table}[!htbp]
  \caption{RSL field design decisions and theoretical grounding.}
  \label{tab:rsl-rationale}
  \small
  \begin{tabular}{@{} l p{4.0cm} l @{}}
    \toprule
    \textbf{Field} & \textbf{Design Decision} & \textbf{Basis} \\
    \midrule
    \texttt{dialogue} & Verbatim from chat log; locked & DP1\textsuperscript{a} \\
    \texttt{narrator} & Factual context for each beat & DP1 \\
    \texttt{intensity} & Normalized [0,1]; drives visual cues & Conflict arc\textsuperscript{b} \\
    \texttt{beat count} & Dynamic 5--12, depending on chat length & Pilot testing\textsuperscript{c} \\
    \texttt{thoughts.text} & AI-generated; accept/modify/reject & DP1 \\
    \texttt{emotion} & Emotion tag with pose and icon cues & DP3\textsuperscript{d} \\
    \texttt{spatial} & Position, pose, and facing per beat & DP3 \\
    \texttt{proxemic} & Relational distance state per beat & \cite{hall1966hidden} \\
    \bottomrule
  \end{tabular}

  \raggedright\footnotesize\vspace{4pt}
  \textsuperscript{a}Dialogue immutability preserves a factual anchor so that reflection operates on interpretations rather than contested wording.\\
  \textsuperscript{b}The sequence models escalation and de-escalation across a conflict arc~\cite{gottman1999seven}.\\
  \textsuperscript{c}Pilot testing informed this range by balancing coverage of the conflict arc with annotation burden.\\
  \textsuperscript{d}Discrete visual cues preserve theatrical abstraction while conveying emotional quality.
\end{table}

% \subsection{Example Beat}

% The following illustrative high-intensity beat shows the RSL structure:

% \begin{Verbatim}[fontsize=\small, frame=single]
% {
%   id: 4,
%   intensity: 0.85,
%   narrator: "B replies after A asks for his attention.",
%   dialogue: {
%     speaker: "B",
%     text: "I said I'm busy! Can you stop
%            doing this every time!"
%   },
%   thoughts: {
%     A: {
%       text: "He's yelling at me again.
%              I just want him to care...",
%       emotion: "hurt"
%     },
%     B: {
%       text: "She doesn't understand my
%              pressure. Why is it always
%              like this!",
%       emotion: "angry"
%     }
%   },
%   spatial: {
%     A: {x:42, facing:"right", pose:"hurt", scale:1.0},
%     B: {x:60, facing:"left", pose:"angry", scale:1.1}
%   },
%   proxemic: { state: "hot", divider: true }
% }
% \end{Verbatim}

\section{Scenario Generation}
\label{app:prompts}
\revthree{
ASIDE uses Gemini~2.5~Flash to transform the complete chat log and contextual input from both partners into a structured RSL scenario. The model selects representative conversational beats and generates revisable inner-state hypotheses, factual narration, emotion labels, theatrical staging, and a context-matched scene preset. It does not generate or modify the original dialogue.}

\subsection{Model Input}
\revthreebody{
Before entering the main workflow, each partner independently reports their central concern, dominant emotion, and relevant situational context. The model receives these responses together with the complete chat log. The contextual input helps interpret what may have remained unspoken, while the chat record remains the factual basis for beat selection and reconstruction.}

\subsection{Abridged Prompt}

The following excerpt preserves the principal instructions and constraints used to generate the study scenarios.

\begingroup
\ifdefined\highlightedcopy\fi
\begin{quote}
\small

You are the inner-state inference and scene reconstruction engine for ASIDE, a dyadic conflict reflection system.

Given a real text-based conflict, the complete chat record, and contextual input from both partners, construct a structured theatrical scenario.

Select 5--12 representative conversational beats that preserve the conflict's progression. Prioritize moments involving escalation, misunderstanding, attempted repair, withdrawal, or changes in emotional direction.

For each beat:

\begin{enumerate}
  \item Copy the selected dialogue exactly from the original chat record. Do not rewrite, summarize, paraphrase, or translate it.
  \item Generate a tentative first-person inner-state hypothesis for each partner. These hypotheses must remain revisable by the participants.
  \item Assign an emotion label and specify each character's pose, position, facing direction, scale, and relational distance.
  \item Provide a concise narrator line describing timing, conversational progression, or observable action. Do not infer emotion in the narrator text.
\end{enumerate}

Select a scene preset that matches the conversational context. Match the tone and intensity of the original exchange. Do not dramatize, exaggerate, invent events, or add unsupported emotional weight. Generate inner-state hypotheses in natural conversational Chinese.

Return only a valid JSON object conforming to the RSL schema.

\end{quote}
\endgroup

\subsection{Prompt Constraints}

The generation prompt imposed the following constraints:

\begin{enumerate}
  \item \textbf{Dialogue fidelity.}
  All selected dialogue must be copied verbatim from the chat record and cannot be generated, translated, or altered by the model.

  \item \textbf{Beat selection.}
  The model selects 5--12 beats according to the length and structure of the conflict. It prioritizes escalation points, misunderstandings, failed repair attempts, withdrawal, and silence gaps while compressing repeated or purely factual exchanges.

  \item \textbf{Tentative inference.}
  Inner states are written as first-person hypotheses rather than factual claims about what either partner actually felt. Participants can accept, revise, or reject them during annotation.

  \item \textbf{Tone control.}
  The generated hypotheses must remain consistent with the original exchange and avoid unnecessary dramatization, accusation, or emotional amplification.

  \item \textbf{Factual narration.}
  Narrator text may describe timing, conversational progression, or observable behavior but cannot assign motives or emotions to either partner.

  \item \textbf{Structured output.}
  The response must be valid JSON conforming to the RSL schema in Appendix~\ref{app:rsl}.
\end{enumerate}

\subsection{Generation Settings and Verification}

Generation used Gemini~2.5~Flash with a temperature of 0.75 and a maximum output length of 12,000 tokens. Before annotation, the research team verified dialogue fidelity, RSL-conformant JSON, and the absence of unsupported events or substantial tone amplification. Outputs that failed these checks were regenerated. All inner states remained revisable hypotheses rather than factual accounts.

\section{Theater Rendering Layers}
\label{app:rendering}

The frontend renders each RSL scenario as a pixel-art theatrical scene. It loads a pre-generated background from a library of 25 presets, selected to match the chat context, and composes the remaining elements at runtime.

\begin{enumerate}
  \item \textbf{Background and scene elements.} The selected background establishes a contextual setting. Pre-made furniture and props add environmental detail.
  \item \textbf{Characters.} Pixel-art sprites render the pose, facing direction, position, and scale specified for each beat. Partner color coding remains consistent across characters and interface elements.
  \item \textbf{Thought bubbles and dialogue.} Thought bubbles display revisable inner-state hypotheses in each partner's color, with an emotion tag and icon; dialogue boxes display the locked text from \texttt{dialogue.text}.
  \item \textbf{Tension and framing cues.} Ambient particles, relational spacing, a proxemic divider when the beat specifies one, and cinematic overlays communicate the scene's theatrical framing and tension.
\end{enumerate}

\section{Synchronization Protocol}
\label{app:sync}

ASIDE uses a WebSocket-based server to manage one private room for each couple. Each room connects two browser clients, identified as Partner A and Partner B. The server maintains their contextual input, cross-edits, self-confirmed annotations, synchronization states, and behavioral logs.

% \subsection{Phase Gate}

% Each room maintains a readiness field for each client. When a partner completes an editing phase, the client sends \texttt{sync:phase\_ready} with the target phase. The server records this value, keeps the stored annotations hidden, and informs the other client that its partner is ready. Once both clients report readiness for the same target phase, the server resets the readiness fields and sends \texttt{sync:phase\_go} to both clients.

% At the Together Viewing gate, this message also carries the other partner's self-confirmations and cross-edits. Both clients therefore enter the phase with complete annotation sets in the same synchronized transition. If a client reconnects while waiting, it resends its readiness state and annotation payload to restore the gate.
\subsection{Phase Gate}

% The phase gate preserves independent editing until both partners are ready for synchronized reveal: (1) Partner A completes an editing phase and sends \texttt{sync:phase\_ready} with the target phase. A then enters a waiting state, while A's annotations remain hidden from B's interface.
% (2) The server records A's readiness and notifies Partner B through \texttt{sync:phase\_partner\_ready}. B can see that A is ready but cannot see A's edits.
% (3) Partner B completes the same phase and sends \texttt{sync:phase\_ready}. Until this point, neither partner can inspect the other's annotations.
% (4) Once both clients report readiness for the same target phase, the server resets the readiness fields and sends \texttt{sync:phase\_go} to both clients. At the Together Viewing gate, this message also carries each partner's self-confirmations and cross-edits to the other client.
% (5) Both clients enter Together Viewing in the same synchronized transition and display the completed annotation sets. Subsequent beat navigation is synchronized so that both partners view the same conversational moment.
% If a client reconnects while waiting, it resends its readiness state and annotation payload so that the server can restore the gate without revealing either partner's annotations prematurely.
The phase gate preserves independent editing until both partners are ready for synchronized reveal: 
\begin{enumerate}
  \item Partner A completes an editing phase and sends \newline
  \texttt{sync:phase\_ready} with the target phase. A then enters a waiting state, while A's annotations remain hidden from B's interface.

  \item The server records A's readiness and notifies Partner B through \texttt{sync:phase\_partner\_ready}. B can see that A is ready but cannot see A's edits.

  \item Partner B completes the same phase and sends \newline
  \texttt{sync:phase\_ready}. Until this point, neither partner can inspect the other's annotations.

  \item Once both clients report readiness for the same target phase, the server resets the readiness fields and sends \texttt{sync:phase\_go} to both clients. At the Together Viewing gate, this message also carries each partner's self-confirmations and cross-edits to the other client.

  \item Both clients enter Together Viewing in the same synchronized transition and display the completed annotation sets. Subsequent beat navigation is synchronized so that both partners view the same conversational moment.
\end{enumerate}
If a client reconnects while waiting, it resends its readiness state and annotation payload so that the server can restore the gate without revealing either partner's annotations prematurely.

\subsection{Message Protocol}

Table~\ref{tab:ws-protocol} summarizes the WebSocket messages used to establish couple rooms, coordinate phase transitions, exchange annotations, synchronize playback, and collect behavioral logs.

% \subsection{Phase Gate}

% Each WebSocket room maintains a readiness field for each client.
% When a partner completes an editing phase and confirms that the current annotations are ready, the client sends \texttt{sync:phase\_ready} with the target phase.
% The server records this value but continues to withhold the stored annotations. It notifies the other client with \texttt{sync:phase\_partner\_ready}. Once both readiness fields contain the same target phase, the server resets them and broadcasts a transition message to both clients (\texttt{sync:phase\_go}). At the Together Viewing gate, this message also carries the other partner's stored self-confirmations and cross-edits. Both clients therefore enter the phase with the completed annotation sets. After reconnecting, a waiting client resends its readiness and annotation payloads to restore the gate.

\begin{table}[h]
  \ifdefined\highlightedcopy\color{ReviewerThree}\fi
  \caption{WebSocket message protocol. C denotes a client, S denotes the server, and Both denotes both clients. $C\rightarrow S$ indicates a client-to-server message, $S\rightarrow C$ a server-to-client message, $S\rightarrow\mathrm{Both}$ a broadcast to both clients, and $C_1\rightarrow S\rightarrow C_2$ a message relayed from one client to the other through the server.}
  \label{tab:ws-protocol}
  \small
  \begin{tabular}{@{} l l l @{}}
    \toprule
    \textbf{Category} & \textbf{Message} & \textbf{Direction} \\
    \midrule
    Room       & \texttt{room:create}                & $C\rightarrow S$ \\
               & \texttt{room:join}                  & $C\rightarrow S$ \\
               & \texttt{room:created}               & $S\rightarrow C$ \\
               & \texttt{room:joined}                & $S\rightarrow C$ \\
               & \texttt{room:partner\_connected}    & $S\rightarrow C$ \\
    \midrule
    Input      & \texttt{input:submit}               & $C\rightarrow S$ \\
               & \texttt{input:partner\_ready}       & $S\rightarrow C$ \\
               & \texttt{input:both\_ready}          & $S\rightarrow\mathrm{Both}$ \\
    \midrule
    Phase gate & \texttt{sync:phase\_ready}          & $C\rightarrow S$ \\
               & \texttt{sync:phase\_partner\_ready} & $S\rightarrow C$ \\
               & \texttt{sync:phase\_go}             & $S\rightarrow\mathrm{Both}$ \\
    \midrule
    Annotation & \texttt{annotation:update}          & $C\rightarrow S$ \\
               & \texttt{annotation:reveal}          & $C_1\rightarrow S\rightarrow C_2$ \\
    \midrule
    Playback   & \texttt{sync:beat}                  & $C_1\rightarrow S\rightarrow C_2$ \\
               & \texttt{sync:phase}                 & $C_1\rightarrow S\rightarrow C_2$ \\
    \midrule
    Logging    & \texttt{log:event}                  & $C\rightarrow S$ \\
    \bottomrule
  \end{tabular}
\end{table}

\subsection{Annotation Exchange}

During Cross-Editing, each \texttt{annotation:update} message updates the sending partner's server-side record but is not forwarded to the other client. Self-confirmed annotations likewise remain private during editing. When a partner finishes, the client sends an \texttt{annotation:reveal} payload containing its self-confirmations and interpretations of the other partner. The server stores this payload and relays it to the other client, which receives the data without displaying it before the phase gate opens. The same annotation sets are included in \texttt{sync:phase\_go} when both partners enter Together Viewing. This second delivery path ensures that both clients receive complete annotation sets even if an earlier reveal message was interrupted.

After the exchange, each character's thought bubble in the theatrical replay displays the other partner's edited interpretation of that character's inner state. When an AI-generated hypothesis was revised, the original text remains visible in struck-through form beneath the edited version. The theatrical replay does not place self-confirmed states beside partner interpretations. That comparison appears in the Divergence Cards, which pair each person's self-confirmed account with the other partner's interpretation at the same conversational beat.

\subsection{Event Logging}

The clients send behavioral events to the server through \texttt{log:event}. The server adds a timestamp, participant role, phase, and room identifier before appending each event to a JSONL file. This append-only record preserves events as they occur. When both clients disconnect, the server also saves a full JSON file containing the session metadata and complete event sequence. Table~\ref{tab:log-events} summarizes the event categories used in the study.

\begin{table}[htbp]
  \ifdefined\highlightedcopy\color{ReviewerThree}\fi
  \caption{Behavioral event categories logged by ASIDE.}
  \label{tab:log-events}
  \footnotesize
  \setlength{\tabcolsep}{2pt}
  \begin{tabular}{@{} p{1.05cm} p{3.55cm} p{3.05cm} @{}}
    \toprule
    \textbf{Category} & \textbf{Events} & \textbf{Recorded data} \\
    \midrule
    Annotation
      & \texttt{self\_confirm}, \texttt{assumption\_confirm}, \texttt{assumption\_edit}, \texttt{assumption\_dispute}, \texttt{assumption\_clear}
      & Beat and persona IDs, original and edited text, status, and emotion changes \\
    Phase
      & \texttt{phase\_change}, \texttt{self\_confirm\_finished}, \texttt{annotation\_finished}
      & From/to phase, completion time, and annotation counts \\
    Playback
      & \texttt{playback\_play}, \texttt{playback\_pause}, \texttt{beat\_advance}, \texttt{beat\_seek}
      & Beat index, navigation trigger, timestamp, and derived dwell duration \\
    Visibility
      & \texttt{bubble\_visibility\_toggle}
      & Visibility state and timestamp \\
    Session
      & \texttt{session\_start}, \texttt{room\_create}, \texttt{room\_join}, \texttt{partner\_connected}, \texttt{partner\_disconnected}
      & Session and scenario IDs, role, room state, and connection changes \\
    \bottomrule
  \end{tabular}
\end{table}

\section{Study Materials}

\subsection{Questionnaire Items}
\label{app:questionnaire}

All questionnaire items used a 7-point Likert scale ranging from 1 (\emph{Strongly Disagree}) to 7 (\emph{Strongly Agree}). S1--S4 were administered at both PRE and POST, whereas S5 was administered only at POST.

\paragraph{Perspective-Taking (S1, four items).}
Adapted from the Interpersonal Reactivity Index Perspective-Taking subscale~\cite{davis1983} and modified to be conflict-specific.
\begin{enumerate}
  \item Before criticizing my partner, I try to imagine how I would feel if I were in their place.
  \item I sometimes try to understand my partner by considering their perspective.
  \item When I am upset with my partner, I usually try to put myself in their shoes.
  \item I find it difficult to see things from my partner's point of view. \emph{(Reverse-scored)}
\end{enumerate}

\textit{Perspective Confidence} (S2, three items). We developed these items based on research on empathic accuracy and perspective mistaking to assess participants' confidence in understanding their partner's thoughts, feelings, and perspective~\cite{ickes1993empathic,eyal2018mind}.

\begin{enumerate}
  \item I have a good understanding of what my partner was truly thinking during that conflict.
  \item I can accurately assess my partner's emotional state at the time.
  \item I clearly understand why my partner acted as they did during that conflict.
\end{enumerate}

\paragraph{Conflict Engagement (S3, four items).}
The items were informed by reflection and behavioral-approach measures~\cite{trapnell1999,elliot2006hierarchical} and tailored to the focal conflict.
\begin{enumerate}
  \item I am willing to carefully review the details of that conflict.
  \item I want to understand more deeply what happened during that conflict.
  \item I want to have a good conversation with my partner about that conflict.
  \item I believe that conflict can be understood and resolved through dialogue.
\end{enumerate}

\paragraph{Empathic Concern (S4, three items).}
The items were informed by the Interpersonal Reactivity Index Empathic Concern subscale~\cite{davis1983} and tailored to the focal conflict.
\begin{enumerate}
  \item When I learn what my partner went through during that conflict, I feel distressed.
  \item Seeing that my partner was hurt during that conflict makes me feel concerned for them.
  \item I can feel my partner's emotions during that conflict.
\end{enumerate}

\paragraph{Scaffolding Effectiveness (S5, POST-only, three items).}
Self-designed items based on scaffolding functions~\cite{wood1976tutoring}.
\begin{enumerate}
  \item The system's initial descriptions helped me begin thinking about my partner's feelings.
  \item Editing existing descriptions made it easier to express my own thoughts than writing from scratch.
  \item Even when some of the system's descriptions differed from what actually happened, overall they still helped me understand my partner more deeply.
\end{enumerate}

\subsection{Interview Protocol}
\label{app:interview}

The questions below formed a semi-structured guide rather than a fixed script. The interviewer adapted their order and wording to the flow of each session, used follow-up probes to clarify or deepen participants' responses, and omitted questions that participants had already answered during the preceding discussion. All interviews nevertheless covered the core topics represented below.

\paragraph{Joint interview questions.}
\begin{enumerate}
  \item When the AI inferred your partner's thoughts, how did you decide whether to edit them? (RQ1)
  \item During Together Viewing, was there a moment that surprised you? (RQ2)
  \item Has your understanding of your partner changed? What contributed to that change? (RQ2)
  \item How did viewing the conflict as a pixel-art replay differ from re-reading the chat log? (RQ2)
  \item Were the AI's initial descriptions helpful, or did they interfere with your thinking? (RQ1)
  \item Was there a moment when you especially wanted to pause and talk to your partner? (RQ2)
\end{enumerate}

\paragraph{Individual follow-up questions.}
\begin{enumerate}
  \item Was there anything you felt but did not say in front of your partner? (RQ1)
  \item After this experience, do you feel your mutual understanding has changed? (RQ2)
  \item Would you want to use the system after a future conflict? Why or why not? (RQ2)
  \item Was there a system description that felt completely wrong or implausible? What was your first reaction? (RQ1)
\end{enumerate}

\end{document}